\documentclass[11pt]{article}

\usepackage[dvips]{graphicx}

\usepackage{amsmath,amssymb}
\usepackage{cite}
\usepackage{comment}

\usepackage{type1cm}
\allowdisplaybreaks[1]

\newcommand{\nt}{\nonumber\\}
\newcommand{\nn}{\nonumber}

\newcommand{\tx}{\text}
\newcommand{\eps}{\epsilon}
\newcommand{\Ups}{\Upsilon}
\newcommand{\emp}{\emptyset}
\newcommand{\Gam}{\Gamma}
\newcommand{\alp}{\alpha}
\newcommand{\gam}{\gamma}

\newcommand{\lag}{\langle}
\newcommand{\rag}{\rangle}
\newcommand{\blag}{\bigl\langle}
\newcommand{\brag}{\bigr\rangle}
\newcommand{\qqquad}{\qquad\qquad}
\newcommand{\qqqquad}{\qqquad\qqquad}

\newcommand{\back}{\!\!\!\!\!\!}

\newcommand{\bp}{{\bar p}}

\newcommand{\bmu}{{\bar\mu}}

\newcommand{\cN}{{\cal N}}
\newcommand{\cL}{{\cal L}}

\newcommand{\cO}{{\cal O}}

\newcommand{\vha}{{\vec{\hat a}}}
\newcommand{\vhb}{{\vec{\hat b}}}

\newcommand{\ha}{{\hat a}}
\newcommand{\hb}{{\hat b}}

\newcommand{\va}{{\vec a}}
\newcommand{\vb}{{\vec b}}
\newcommand{\ve}{{\vec e}}

\newcommand{\vW}{{\vec W}}
\newcommand{\vY}{{\vec Y}}
\newcommand{\valp}{{\vec \alpha}}
\newcommand{\vbet}{{\vec \beta}}

\newcommand{\vvphi}{{\vec \varphi}}
\newcommand{\vrho}{{\vec \rho}}
\newcommand{\vlam}{{\vec \lambda}}

\newcommand{\Dinf}{\Delta_\infty}
\newcommand{\Di}{\Delta_1}
\newcommand{\Do}{\Delta_0}
\newcommand{\winf}{w_\infty}
\newcommand{\wi}{w_1}
\newcommand{\wo}{w_0}

\newcommand{\udl}{\underline}

\newcommand{\ba}{\begin{eqnarray}}
\newcommand{\ea}{\end{eqnarray}}
\newcommand{\ban}{\begin{eqnarray*}}
\newcommand{\ean}{\end{eqnarray*}}
\newcommand{\sss}[1]{\subsubsection*{#1}}
\newcommand{\sssudl}[1]{\subsubsection*{\udl{\textmd{#1}}}}

\def\Up{\Upsilon}
\def\G2{\Gamma_2}

\def\tV{\tilde V}

\begin{document}
\begin{titlepage}

\begin{flushright}
UT-10-06\\
KEK-TH-1360
\end{flushright}

\vskip 12mm

\begin{center}

{\bfseries\fontsize{15.4pt}{20pt}\selectfont
Analysis of correlation functions in Toda theory\\[+7pt]
and AGT-W relation for SU(3) quiver}

\vskip 12mm
Shoichi Kanno$^\dagger$\footnote{
E-mail address: kanno@hep-th.phys.s.u-tokyo.ac.jp},
{\large
Yutaka Matsuo$^\dagger$\footnote{
E-mail address: matsuo@phys.s.u-tokyo.ac.jp}
and
Shotaro Shiba$^\ddagger$\footnote{
E-mail address: sshiba@post.kek.jp}
 }\\
\vskip 5mm
{\it\large
$^\dagger$
Department of Physics, Faculty of Science, University of Tokyo,\\
Hongo 7-3-1, Bunkyo-ku, Tokyo 113-0033, Japan\\
}
\vskip 4mm
{\it\large
$^\ddagger$
Institute of Particle and Nuclear Studies,\\
High Energy Accelerator Research Organization (KEK),\\
Oho 1-1, Tsukuba-city, Ibaraki 305-0801, Japan\\
\noindent{ \smallskip }\\
}
\vspace{45pt}
\end{center}
\begin{abstract}
We give some evidences of the AGT-W relation between
$SU(3)$ quiver gauge theories and $A_2$ Toda theory.
In particular, we derive the explicit form of 5-point correlation functions
in the lower orders and confirm the agreement with Nekrasov's partition
function for $SU(3)\times SU(3)$ quiver gauge theory.
The algorithm to derive the
correlation functions can be applied to general $n$-point function
in $A_2$ Toda theory which will be useful to establish the relation
for more generic quivers.
Partial analysis is also given for $SU(3)\times SU(2)$
case and we comment on some technical issues
which need clarification before establishing the relation.
\end{abstract}

\end{titlepage}

\setcounter{footnote}{0}

\section{Introduction}
It is well-known, after the seminal works by
Seiberg-Witten \cite{Seiberg:1994rs,Seiberg:1994aj},
that there is a close relation between 4-dim $\cN=2$ gauge theories
and the quantum geometry of 2-dim Riemann surface.
Recently, Gaiotto \cite{Gaiotto:2009we} invented a new representation
of the Seiberg-Witten curve where the duality transformation of
the couplings of $\cN=2$ system
is encoded as the duality transformation of the moduli of the curve.
More precisely, the hypermultiplet in the $\cN=2$ gauge theory is represented
as a puncture on the curve and the gauge field 
is given as the cylinder which connects the punctures.
The gauge coupling is then extracted as the modulus of the cylinder.

The relation between Gaiotto's
curve and $\cN=2$ gauge theories was further deepened by
Alday, Gaiotto and Tachikawa~\cite{Alday:2009aq}.
They observed that the Nekrasov's formula \cite{b:Nekrasov} for the partition function of
$\cN=2$ gauge theory coincides with
the correlation function 
of Liouville field theory.
They obtained such result for $N_f=4$ $SU(2)$ gauge theory
(4-point function on sphere in Liouville side)
 and $\cN=2^*$ $SU(2)$ gauge theory (1-point function on torus),
and conjectured that such relation, which is called  ``AGT relation",
exists for other superconformal field theories.
After this work, many studies have been carried out~\cite{Marshakov:2009gs,Nanopoulos:2009au,Marshakov:2009kj,Mironov:2009qn,Mironov:2009uv,Fateev:2009aw,Giribet:2009hm,Alba:2009ya,Mironov:2010zs,Gaiotto:2009ma,Marshakov:2009gn,Alba:2009fp,Hadasz:2010xp,Poghossian:2009mk},
which analytically prove this relation in some cases or limits.
This relation also have been studied energetically in the context of Dijkgraaf-Vafa matrix model~\cite{Dijkgraaf:2009pc,Itoyama:2009sc,Eguchi:2009gf,Schiappa:2009cc,Mironov:2009ib,Fujita:2009gf,Sulkowski:2009ne,Itoyama:2010ki,Mironov:2010ym,Morozov:2010cq,Eguchi:2010rf}.
Moreover, the loop and surface operators in 4-dim gauge theory and their correspondence in Liouville theory are widely discussed~\cite{Drukker:2009tz,Alday:2009fs,Drukker:2009id,Gaiotto:2009fs,Wu:2009tq,Passerini:2010pr,Drukker:2010jp,Kozcaz:2010af,Alday:2010vg,Gaiotto:2010be,Petkova:2009pe}.
The 5-dim extension of this relation also has been studied~\cite{Awata:2009}.

A natural generalization of AGT relation is the
similar correspondence between  $\cN=2$ $SU(N)$ quiver gauge theories and
$A_{N-1}$ Toda theory~\cite{Wyllard:2009hg}
(``AGT-W relation").
For $SU(N)$ case, the puncture in Gaiotto curve
has an extra label of Young diagram  \cite{Gaiotto:2009we}.
For the linear quiver, two punctures with the general Young diagram
appear on two edges, while the other punctures are labeled
by a special Young diagram $[N-1,1]$.  The latter ones are
 called  ``simple punctures."
In \cite{Wyllard:2009hg}, by using the example of
$SU(N)$ gauge theory with $N_f=2N$, it was conjectured
that the simple puncture is associated with
the level-1 singular vector of $W_N$ algebra, the symmetry
of $A_{N-1}$ Toda field theory. This AGT-W relation is explicitly checked in $SU(3)$ case 
up to instanton number 2 \cite{Mironov:2009by,Taki:2009zd}.
Some analysis for proof in $SU(N)$ case also has been done~\cite{Mironov:2009qt,Mironov:2009dv,Nanopoulos:2010zb,Nanopoulos:2010ga}.
In \cite{Kanno:2009ga}, we generalized this conjecture to the
puncture with general Young diagram and determined the possible
form of the associated vertex operator.  In particular, we
confirmed that Gaiotto's curve can be reproduced
through the null state conditions which the vertex
operators satisfy in the semi-classical limit.

To establish AGT-W relation with such general punctures,
we need develop a method to compute corresponding
correlation functions of Toda theory, at least the 5-point functions,
which are written in the form of Selberg integral.  To obtain
an analytic formula to carry out such integration would be
highly desirable but at this moment it is technically difficult.
As the intermediate step, we will establish, in this paper, an algorithm
where partial result (first few terms in the expansion of moduli parameters)
can be obtained by  computer.
For this purpose, we decompose the Riemann surface
into the ``propagators'' and ``vertices''
as in the perturbative string theory.
In particular, for the family of theories which
are called linear quiver gauge theory, one needs consider only
Riemann sphere as the tree diagrams.

One technical non-triviality
in obtaining 3-point vertex for Toda theory (or $W$-algebra) is
the recursion formula for $W$-generators.
As discussed in the literature, the conformal Ward identity
reduces the number of generators in the correlator
through the highest weight conditions.  For Toda theory,
there remain product of $W_{-1}$ generators which cannot
be simplified further by the recursion formula.
It implies that we need some constraints on the
primary fields in the correlator to solve it.
As shown in \cite{Fateev:2007ab}, a solution is to impose
one of the primary fields in the 3-point function to have
the level-1 null state condition by which one can replace
$W_{-1}$ by $L_{-1}$.
For the linear quiver gauge theory, fortunately, if we decompose
Gaiotto curve into 3-point functions, we have at least one
simple vertex which is indeed characterized by such condition.

We organized the paper as follows.  In section \ref{s:Nek},
we summarize Nekrasov's partition function for linear quiver theory
which should be reproduced as the correlation function of Toda theory.
In section \ref{sec:corr}, we review the known formulae on
the correlation functions of Toda field theory after \cite{Fateev:2007ab}.
In particular, the information of the coefficient of 3-point function is
essential to derive the 1-loop part of partition functions.  We also review the
general strategy of the computation of the correlation functions by
decomposing the curve into the ``propagators'' and ``vertices.''

In section \ref{sec:3pt}, we explain the explicit algorithm
of the calculation of the conformal blocks in the lower orders.
A nontrivial part of this section is the derivation of recursion
formula for the 3-point function.  As we noted, imposing
the level-1 null state condition for one of the operators
in the 3-point function is essential to fix their explicit form.

In section \ref{sec:sol}, we apply our general strategy to
a specific quiver gauge theory, $SU(3)\times SU(3)$ quiver,
which corresponds to a 5-point function in Toda theory.
We have confirmed the correspondence in both 1-loop
and the instanton contribution at the lower levels.
We also work on $SU(3)\times SU(2)$ case.
This time, however, we meet some problems. A naive calculation
of the 5 point function shows that it vanishes automatically.
In order to proceed further, we removing the factor which vanishes by hand.
After that, the remaining factors reproduce some part of the
Nekrasov's partition function.  However,
the correspondence is not complete yet.
We argue that we need to tune some parameters (such as the mass of the
bifundamental matter field) in order to meet the gauge side.
We doubt that the correlation function of Toda theory may need
some modification when two of the vertex operators in 3-point function
have null states at level one.  We give some preliminary analysis in
appendix \ref{s:3pt}.
We hope to clarify this in the future work.

\section{AGT-W relation}

AGT relation~\cite{Alday:2009aq}
reveals the nontrivial correspondence between
the partition function of 4-dim $\cN=2$ $SU(2)$ quiver gauge
theory and
the correlation function of 2-dim Liouville (or $A_1$ Toda)
field theory.
The 2-dim theory is defined on Seiberg-Witten curve
which determines the field contents of the 4-dim theory.
The Seiberg-Witten system can be interpreted as
the intersecting D4/NS5-branes' system, where the intersection points
exist on Seiberg-Witten curve \cite{Gaiotto:2009we,Witten:1997sc}.
AGT relation says that
when we insert the Liouville vertex operators at all the intersection points and calculate the correlation function of these vertex operators in the 2-dim theory,
this function can be rewritten as the product of the partition function of the 4-dim theory and some additional factors.
Schematically, it is written as
\ba\label{AGTcorr}
\left|Z^{\mathrm{gauge}}\right|^2
=
\langle V\cdots V\rangle^{\mathrm{Liouville}}
\ea
where $Z^{\mathrm{gauge}}$ is the partition function of
quiver gauge theory.
In the right hand side, we have the conformal block of Liouville theory
where $V$ is the vertex operator insertion.

A natural generalization of AGT relation seems
a similar correspondence between the partition function
of 4-dim $\cN=2$ $SU(N)$ quiver gauge theory and
the correlation function of 2-dim $A_{N-1}$ Toda field
theory~\cite{Wyllard:2009hg}.
In this paper, we examine this conjecture for $N=3$ case,
that is, the case of $SU(3)$ quiver gauge theory and
$A_2$ Toda field theory.
First, in this section, we review the form of the functions
on the both sides.

\subsection{4-dim gauge theory side: partition functions}
\label{s:Nek}

\begin{figure}[bpt]
\begin{center}
\includegraphics[scale=0.6]{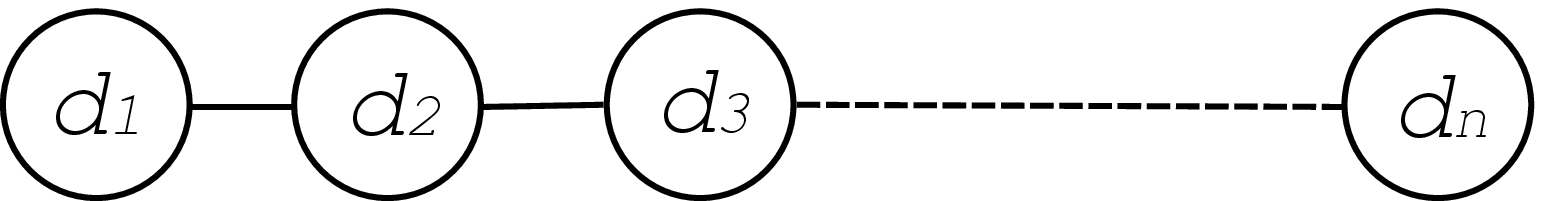}
\end{center}
\caption{Linear quiver gauge theory}
\label{fig:quiver}
\end{figure}

The partition function of 4-dim $\cN=2$ gauge theory is
written as the product of classical part, 1-loop correction part and
nonperturbative instanton correction part:
\ba
Z^{\mathrm{gauge}}=Z_\text{class}\, Z_\text{1-loop}\, Z_\text{inst}\,.
\ea
In the following, we consider the case of
$\prod_{k=1}^n SU(d_k)$ linear quiver gauge theory (fig.\,\ref{fig:quiver}).
Especially, we will consider only the conformal invariant cases where
$k_a= 2d_a-d_{a+1}-d_{a-1}(\geq 0)$ fundamental hypermultiplets are attached to gauge group $SU(d_a)$.

The classical part of the partition function is
\ba
Z_\tx{class}=\exp\left[\sum_{k=1}^n 2\pi i \tau_k |\vec{\hat a}_k|^2\right],
\ea
where $\tau_k:=\frac{\theta_k}{2\pi}+\frac{4\pi i}{g_k^2}$ is
the complex UV coupling constant.
$\vec{\hat a}_k:=\sum_{i=1}^{d_k-1}a_i\vec{e}_i$ is
the diagonal of VEV's $a_i$ of adjoint scalars,
where $\vec e_i$ are the simple roots of gauge symmetry algebra.
For $SU(d)$ algebra, we usually set
\ba\label{root}
\ve_i=(0,\cdots,0,1,-1,0,\cdots,0)\,,
\ea
where $1$ is $i$-th element. It gives, for example,
$\vec{\hat a}=(a_1,-a_1)$ for $SU(2)$ and
$\vec{\hat a}=(a_1,-a_1+a_2,-a_2)$ for $SU(3)$.

The 1-loop contribution to the partition function
is~\cite{Alday:2009aq,Wyllard:2009hg}
\ba\label{Z1lp}
Z_\text{1-loop}
&=&
 \left(\prod_{k=1}^n z^\tx{1lp}_\tx{vec}(\va_k)\right)
 \left(\prod_{\bp=1}^{d_1} z^\tx{1lp}_\tx{afd}(\va_1,\bmu_\bp)\right)
\nt&&\times
 \left(\prod_{k=1}^{n-1} z^\tx{1lp}_\tx{bfd}(\va_k,\va_{k+1},m_k)\right)
 \left(\prod_{p=1}^{d_n} z^\tx{1lp}_\tx{fd}(\va_n,\mu_p)\right),
\ea
where
$\mu_p, \bar\mu_\bp, m_k$ are the mass of fundamental, antifundamental, bifundamental fields, respectively, and
\ba\label{1lpfactor}
z^\tx{1lp}_\tx{vec}(\va)
&=&
\prod_{i<j}\exp\left[-\gamma_{\eps_1,\eps_2}(\ha_i-\ha_j-\eps_1)
-\gamma_{\eps_1,\eps_2}(\ha_i-\ha_j-\eps_2)\right],\nt
z^\tx{1lp}_\tx{fd}(\va,\mu)
&=&
\prod_{i}\exp\left[\gamma_{\eps_1,\eps_2}(\ha_i-\mu)\right],\nt
z^\tx{1lp}_\tx{afd}(\va,\bar\mu)
&=&
\prod_{i}\exp\left[\gamma_{\eps_1,\eps_2}(-\ha_i+\bar\mu-\eps_+)\right],\nt
z^\tx{1lp}_\tx{bfd}(\va,\vb,m)
&=&
\prod_{i,j}\exp\left[\gamma_{\eps_1,\eps_2}(\ha_i-\hb_j-m)\right],
\ea
where
$\eps_+:=\eps_1+\eps_2$
($\eps_1$, $\eps_2$ are Nekrasov's deformation parameters), and
the function $\gamma_{\eps_1,\eps_2}(x)$ is related to double Gamma
function $\Gamma_2(x|\eps_1,\eps_2)$ as\footnote{
We would like to thank Satoshi Nawata for pointing out typos in the previous version: the definition of $\gamma_{\eps_1,\eps_2}(x)$ in eq.\,(\ref{def-gam}) and the term for $\vlam_i$ just above eq.\,(\ref{vlam}) .}
\ba\label{def-gam}
\gamma_{\eps_1,\eps_2}(x)=\log \Gamma_2(x+\eps_+|\eps_1,\eps_2)\,.
\ea
The properties of double Gamma function are summarized in appendix \ref{sec:app}.

The instanton contribution  is obtained by Nekrasov's instanton counting
formula with Young tableaux~\cite{Alday:2009aq,Alba:2009ya}:
\ba
Z_\tx{inst}
&=&\sum_{\{\vY_1,\cdots,\vY_n\}}
 \left(\prod_{k=1}^n q_k^{|\vY_k|}z_\tx{vec}(\vha_k,\vY_k)\right)
 \left(\prod_{\bp=1}^{d_1} z_\tx{afd}(\vha_1,\vY_1,\bmu_\bp)\right)
\nt&&\times
 \left(\prod_{k=1}^{n-1} z_\tx{bfd}(\vha_k,\vY_k;\vha_{k+1},\vY_{k+1};m_k)\right)
 \left(\prod_{p=1}^{d_n} z_\tx{fd}(\vha_n,\vY_n,\mu_p)\right),
\ea
where
$q_k:=e^{2\pi i\tau_k}$ ($\tau_k$ is the coupling constant),
and
$\vec Y_k=(Y_{k,1},\cdots,Y_{k,d_k})$ is a set of Young tableaux.
$|\vec Y_k|$ is the total sum of number of boxes of Young tableaux
$Y_{k,i}$ ($i=1,\cdots,d_k$).
Each factor of the instanton part is written as
\ba
z_\tx{bfd}(\vha,\vY;\vhb,\vW;m)&=&
 \prod_{i,j} \prod_{s\in Y_i}(E(\ha_i-\hb_j,Y_i,W_j,s)-m)
\nt&&\times
 \prod_{t\in W_j}(\eps_+-E(\hb_j-\ha_i,W_j,Y_i,t)-m)\,,\nt
z_\tx{vec}(\vha,\vY)&=&1/z_\tx{bfd}(\vha,\vY;\vha,\vY;0)\,,\nt
z_\tx{fd}(\vha,\vY,\mu)&=&
 \prod_{i} \prod_{s\in Y_i}(\phi(\ha_i,s)-\mu+\eps_+)\,,\nt
z_\tx{afd}(\vha,\vY,\bmu)&=&z_\tx{fd}(\vha,\vY,\eps_+-\bmu)\,.
\ea
The functions $E(\ha,Y,W,s)$ and $\phi(\ha,s)$ are defined as
\ba
E(\ha,Y,W,s)
&=&\ha-\eps_1(\lambda'_{W,j}-i)+\eps_2(\lambda_{Y,i}-j+1)\,,\nt
\phi(\ha,s)&=&\ha+\eps_1(i-1)+\eps_2(j-1)\,,
\ea
where $s=(i,j)$ denotes the position of the box in a Young tableau
({\em i.e.}~the box in $i$-th column and $j$-th row).
$\lambda_{Y,i}$ is the height of $i$-th column, and
$\lambda'_{Y,j}$ is the length of $j$-th row for Young tableau $Y$.
That is,
$\lambda'_{Y,j}-i$ 
and
$\lambda_{Y,i}-j$ 
are the length of `leg' and `arm' of the Young tableau $Y$
for the box $s=(i,j)$, respectively.

\subsection{2-dim CFT: W-algebra, Toda theory, and correlation functions}
\label{sec:corr}

The action of 2-dim $A_{N-1}$ Toda field theory is
\ba \label{TodaAction}
S=\int d^2\sigma\sqrt{g}\left[
\frac{1}{8\pi}g^{xy}\partial_x\vvphi\cdot\partial_y\vvphi
+\mu\sum_{k=1}^{N-1}e^{b\ve_k\cdot\vvphi}
+\frac{Q}{4\pi}R\vrho\cdot\vvphi
\right],
\ea
where $\vvphi=(\varphi_1,\cdots,\varphi_N)$ is the Toda fields
satisfying $\sum\varphi_k=0$.
$g_{xy}$ is the metric on 2-dim Riemann surface,
and $R$ is its curvature.
$\ve_k$ is the $k$-th simple root written as eq.\,(\ref{root}),
and $\vrho$ is the Weyl vector
({\em i.e.}~half the sum of all positive roots)
of $A_{N-1}$ algebra.
$b$ is a real parameter, and $Q:=b+1/b$.
This theory is conformal invariant with the central charge
\ba
c=(N-1)+12Q^2\vrho\cdot\vrho=(N-1)(1+N(N+1)Q^2)\,.
\ea

The symmetry algebra of this theory is generated by
the energy-momentum tensor $T(z)$ and
additional $N-2$ chiral currents $W^{(3)},\cdots,W^{(N)}$
with spin $3,\cdots,N$.
In the following in this paper, we concentrate on $N=3$ case.
In this case, the following generators are defined as Laurent
expansion of the currents:
\ba\label{TW}
T(z)=:\sum_{n=-\infty}^\infty \frac{L_n}{z^{n+2}}\,,\quad
W^{(3)}(z)=:\sum_{n=-\infty}^\infty \frac{W_n}{z^{n+3}}\,.
\ea
The commutation relation for the generators is given by
\ba\label{comm}
{}[L_n,L_m]&=&(n-m)L_{n+m}+\frac{c}{12}(n^3-n)\delta_{n+m,0}\nt
{}[L_n,W_m]&=&(2n-m)W_{n+m}\nt
{}\frac29[W_n,W_m]&=&\frac{c}{3\cdot 5!}n(n^2-1)(n^2-4)\delta_{n+m,0}
+\frac{16}{22+5c}(n-m)\Lambda_{n+m}
\\&&
+(n-m)\left(\frac{1}{15}(n+m+2)(n+m+3)-\frac16(n+2)(m+2)\right) L_{n+m}
\,,\nn
\ea
where
\ba\label{Lam}
\Lambda_n=\sum_{k=-\infty}^\infty :L_kL_{n-k}:+\frac15 x_nL_n\,,
\ea
with $x_{2l}=(1+l)(1-l)$ and $x_{2l+1}=(2+l)(1-l)$.

In the following, we need the action of such
generators on the operators
at arbitrary point $z=\zeta$. For this purpose, it is useful to introduce
the operators $L_n(\zeta)$ and $W_n(\zeta)$ defined by the contour integration
around $z=\zeta$,
\ba
L_n (\zeta) = \oint_{z=\zeta}\frac{dz}{2\pi i}(z-\zeta)^{n+1}  T(z)\,,\quad
W_n (\zeta) = \oint_{z=\zeta}\frac{dz}{2\pi i}(z-\zeta)^{n+2}  W(z)\,.
\ea
The operators in eq.\,(\ref{TW}) are identical to the
special case $\zeta=0$, namely $L_n=L_n(0)$ and $W_n=W_n(0)$.
The commutation relations among $L_n(\zeta)$ and $W_n(\zeta)$
are identical to eq.\,(\ref{comm}).

The highest weight (ket) state $|\Delta, w\rangle$
and its conjugate (bra) state $\langle \Delta, w|$ is defined by
the conditions:
\ba
&&\back
 L_n |\Delta, w\rangle=0,
\qqquad
 W_n |\Delta, w\rangle=0,\qquad (n>0) \nn\\
&&\back
 L_0 |\Delta, w\rangle=\Delta |\Delta, w\rangle,
\quad
 W_0 |\Delta, w\rangle= w |\Delta, w\rangle, \label{highest}\\
&&\back
 \langle\Delta, w|L_n =0,
\qqquad
 \langle\Delta, w|W_n=0 ,\qquad (n<0)\nn\\
&&\back
 \langle\Delta, w|L_0 =\langle\Delta, w| \Delta,
\quad
 \langle\Delta, w|W_0=\langle\Delta, w| w\,.\label{highest2}
\ea
The adjoint of operators are defined as
\ba\label{adjoint}
L_n^\dagger = L_{-n}, \quad W_n^\dagger = W_{-n}.
\ea

In terms of Toda fields,  the highest weight state is given by the
vertex operator:
\ba
&& V_{\valp}(z)= :e^{\valp\cdot\vvphi(z)}:
\qquad (\valp\in \mathbf{C}^3, \quad\sum_{i=1}^3\alpha_i=0)\,,\nt
&& |V_{\valp}\rangle =  \lim_{z\rightarrow 0} V_{\valp}(z) |0\rangle
\,,\quad
\langle V_{\valp}|=\lim_{z\rightarrow \infty} z^{2\Delta_{\valp}} \langle 0|V_\valp(z)
\,.
\ea
The parameters $\Delta, w$ are related to $\vec\alpha$ as
\ba
\Delta_\valp=\frac12 (2Q\vrho-\valp)\cdot\valp\,,\quad
w_\valp=i\sqrt{\frac{48}{22+5c}}\,\prod_{i=1}^3\,(\valp-Q\vrho)\cdot\vlam_i\,,
\ea
where $\vlam_i$ are the weights of the fundamental representation of $A_2$ Lie algebra:
\ba\label{vlam}
&&\!\!\!
\vlam_1=\frac13(2,-1,-1)\,,\quad
\vlam_2=\vlam_1-\ve_1=\frac13(-1,2,-1)\,,
\nt&&\!\!\!
\vlam_3=\vlam_2-\ve_2=\frac13(-1,-1,2)\,.
\ea
and $\vrho=(1,0,-1)$ is the Weyl vector.

The inner product $\langle V_{\valp}|V_{\valp}\rangle$ vanishes
in general because of the conservation of momentum.
The nontrivial inner product can be taken in the form
$
\langle V_{2Q\vec\rho - \valp}|V_{\valp}\rangle =1.
$

\sss{Correlation functions}
It is known that the form of general 3-point functions of primary fields
is determined by the conformal invariance as~\cite{Fateev:2007ab}
\ba\label{C}
\left\langle
V_{\valp_1}(z_1) V_{\valp_2}(z_2) V_{\valp_3}(z_3)
\right\rangle
=\frac{C(\valp_1,\valp_2,\valp_3)}
{|z_{12}|^{2(\Delta_1+\Delta_2-\Delta_3)}
 |z_{13}|^{2(\Delta_1+\Delta_3-\Delta_2)}
 |z_{23}|^{2(\Delta_2+\Delta_3-\Delta_1)}}\,,\qquad
\ea
where $z_{ij}:=z_i-z_j$ and $\Delta_i:=\Delta_{\valp_i}$.
It is difficult to calculate the coefficients
$C(\valp_1,\valp_2,\valp_3)
=\left\langle V_{\valp_1}|V_{\valp_2}(1) |V_{\valp_3}\right\rangle$
for general $\valp_i$, however, if one of $\valp_i$'s is proportional
to $\vlam_1$ or $-\vlam_3$
(the highest weight of fundamental or antifundamental representation),
they are obtained as
\ba\label{C1}
C(\valp_1,\valp_2,\gamma\vlam_1)
&=&
 \left[\pi\mu\gam(b^2)b^{2-2b^2}\right]^{(2Q\vrho-\sum\valp_i)\cdot\vrho/b}\\
&&\times \frac{\Ups(b)^2 \Ups(\gamma)\prod_{e>0}
 \Ups\bigl((Q\vrho-\valp_1)\cdot\ve\bigr)
 \Ups\bigl((Q\vrho-\valp_2)\cdot\ve\bigr)}
{\prod_{i,j}\Ups\bigl(\frac13\gamma-
  (\valp_1-Q\vrho)\cdot\vlam_i-
  (\valp_2-Q\vrho)\cdot\vlam_j\bigr)}\,,
\nn\ea
and
\ba\label{C2}
C(\valp_1,\valp_2,-\gamma\vlam_3)
&=&
 \left[\pi\mu\gam(b^2)b^{2-2b^2}\right]^{(2Q\vrho-\sum\valp_i)\cdot\vrho/b}\\
&&\times \frac{\Ups(b)^2 \Ups(\gamma)\prod_{e>0}
 \Ups\bigl((Q\vrho-\valp_1)\cdot\ve\bigr)
 \Ups\bigl((Q\vrho-\valp_2)\cdot\ve\bigr)}
{\prod_{i,j}\Ups\bigl(\frac13\gamma+
  (\valp_1-Q\vrho)\cdot\vlam_i+
  (\valp_2-Q\vrho)\cdot\vlam_j\bigr)}\,,
\nn\ea
where $\gamma$ is a constant, and
$\prod_{e>0}$ means the product over all positive roots.
$\Ups$ is the Upsilon function whose properties are summarized in
appendix \ref{sec:app}.

In order to calculate general $n$-point functions, we often use
the following decomposition~\cite{Mironov:2009dr}: 
\ba\label{sew}
&&\back
\blag\cO_n(z_n)\,,\cdots,\cO_3(z_3)\,,\cO_2(z_2)\,,\cO_1(z_1)\brag
\nt&&\back
=\sum_{\valp,Y_1,Y_2}
(z_2-z_1)^{\Delta_V-\Delta_{\cO_2}-\Delta_{\cO_1}}
\blag\cO_n(z_n)\,,\cdots,\cO_3(z_3)\,,\cL_{-Y_1}(z_1)V_\valp(z_1)\brag
\nt&&\qquad\quad\times
(S^{-1}_\valp)_{Y_1,Y_2}
\blag V_{2Q\vec\rho-\valp}\,|(\cL_{-Y_2})^\dagger \,\cO_2(1)|\cO_1\brag\,,
\ea
where the index $\valp$ labels the primary fields
and $Y_1,Y_2$ labels the descendants of the primary fields, such as
\ba
\cL_{-Y}(z) V_\valp(z):=
 L_{-n_1}(z) \cdots L_{-n_l}(z)
 W_{-n'_1}(z) \cdots W_{-n'_w}(z)  V_\valp(z)\,.
\ea
Here $Y$ is a set of two Young tableaux $(Y_L,Y_W)$
with
$Y_L=(n_1\geq \cdots\geq n_l)$ and
$Y_W=(n'_1\geq \cdots\geq n'_w)$.
It means that $|Y|:=\sum_{i=1}^l n_i+\sum_{j=1}^w n'_j$ is the level of the descendant.
The descendants at the same level $|Y|$ can be labeled
by the partitions of integer $|Y|$,
so it is useful to classify them using the sets of Young tableaux $Y$.
The conformal dimension $\Delta_V$ in eq.\,(\ref{sew}) is that of $\cL_{-Y_2}V_\valp$
which is defined in this way.

The matrix $S$ is called as Shapovalov matrix
\ba\label{Shap}
S_{\valp,Y_1,Y_2}:=
\lag V_{2Q\vec\rho-\valp} |(\cL_{-Y_1})^\dagger\, \cL_{-Y_2} |V_\valp\rag
\ea
Once $Y_1, Y_2$ is given, it is possible to determine Shapovalov matrix
by using the commutation relations (\ref{TW}) and the highest weight conditions
(\ref{highest}), (\ref{highest2}).  We note that unless $|Y_1|=|Y_2|$, the inner
product vanishes. It implies that the Shapovalov matrix
is block diagonal and each block is finite size. It helps us to determine
the inverse Shapovalov matrix by restricting the computation to each level.
The explicit form of the Shapovalov matrix for lower levels
is given in appendix \ref{app:Shap} and \cite{Mironov:2009by}.

When we decompose the $n$-point functions for vertex operators
using such rule repeatedly, 
they necessarily become the linear combination of the products of
inverse Shapovalov matrices (or ``propagators'') $S^{-1}$ and
3-point functions (or ``vertices'').  In the decomposition
(\ref{sew}), together with Shapovalov matrix, we have
correlation function of the form
\ba
&&
\Gamma_{\valp_\infty,\valp_1,\valp_0}(Y_\infty, Y_1, Y_0)
\nt&&\qquad
:=~
\lag V_{2Q\vrho-\valp_\infty}| (\cL_{-Y_\infty})^\dagger~
\cL_{-Y_1}(1) V_{\valp_1}(1)~
\cL_{-Y_0}|V_{\valp_0}\rag
\nt&&\qquad
=:~
\lag \cL_{-Y_\infty} V_{2Q\vec\rho-\valp_\infty},
\cL_{-Y_1} V_{\valp_1},
\cL_{-Y_0}V_{\valp_0}\rag\,.\label{3pt}
\ea
We note that in this expression, the operator at infinity
is defined by the adjoint (\ref{adjoint}).
By repeating the decomposition, we are left with
a 3-point function in the leftmost factor.  Here we have to be
careful in the definition of operator
at $z=\infty$.  In order to keep conformal property,
we have to use the conformal transformation, say $w=1/z$,
to define the adjoint (BPZ conjugation).  For Virasoro generators,
the two definitions agree. For $W$-operator, however, BPZ conjugation
($W^\flat$) is given as,
by using $W(z)=W(w) (\frac{dw}{dz})^3=-W(w) w^6$,
\ba
(W_n)^\flat= -W_{-n}= -W_n^\dagger\,.
\ea
Thus we have discrepancy in the sign.  In order to describe
the 3-point function which appears in the leftmost factor,
we need to introduce another type of 3-point function $\Gamma'$:
\ba
&&
\Gamma'_{\valp_\infty,\valp_1,\valp_0}(Y_\infty, Y_1, Y_0)
\nt&&\qquad
:=~
\lag V_{\valp_\infty}| (\cL_{-Y_\infty})^\flat~
\cL_{-Y_1}(1) V_{\valp_1}(1)~
\cL_{-Y_0}|V_{\valp_0}\rag
\nt&&\qquad
=:~
\lag \cL_{-Y_\infty} V_{\valp_\infty},
\cL_{-Y_1} V_{\valp_1},
\cL_{-Y_0}V_{\valp_0}\rag'\,.
\label{3pt'}
\ea
where $\cL_{-Y_\infty}(\infty)$ in the last line
to be defined by BPZ conjugation.
Again, note that $\Gamma'$ is different from $\Gamma$ in the sign.

To summarize, the $n$-point functions can be written in the form:
\ba
\langle \cO_n,\cdots,\cO_3,\cO_2,\cO_1\rangle
\,\sim\,
\sum \Gamma' S^{-1} \Gamma S^{-1} \Gamma \cdots \Gamma S^{-1} \Gamma\,.
\ea
When all the vertices in the correlator $\cO_i$ are primary,
one needs only the special case where $Y_1=\emp$ (no boxes)
in all the 3-point functions $\Gamma$, $\Gamma'$.

If one can determine these 3-point functions,
one can in principle calculate the general
$n$-point correlation functions by patching them together
with the inverse Shapovalov matrices.  As we will see in the next section,
this is indeed possible 
as long as one of the vertex operators in 3-point functions satisfies
$\valp_i \propto \vlam_1$ or $\valp_i \propto -\vlam_3$.
We note that such operators satisfy the null state condition at level-1,
which is necessary to reduce the degree of freedom of $W$-generators.
The above is our strategy to obtain the $n$-point functions.

\section{Recursion formula for 3-point functions}
\label{sec:3pt}

In order to determine the 3-point functions (\ref{3pt}) and (\ref{3pt'}),
we use the deformation of integration contour paths
and derive the recursion formula.
In the remainder of this section,
we write
$\tilde V_\infty:=\cL_{-Y_\infty} V_{\valp_\infty}(\infty)$\,,
$\tilde V_1:=\cL_{-Y_1} V_{\valp_1}(1)$,
$\tilde V_0:=\cL_{-Y_0} V_{\valp_0}(0)$,
for simplicity.

\sss{Recursion formula for $T$ insertion}

First, the standard CFT recursion formula for
$\Gamma$-type 3-point functions (\ref{3pt}) gives
\ba \label{n00}
&&\back
\langle
L_{-n} \tV_\infty,\tV_1,\tV_0
\rangle
=
\oint_\infty \frac{dz}{2\pi i}z^{n+1}
\langle
T(z)\tV_\infty,\tV_1,\tV_0
\rangle
\nt&&\back
=
\oint_1 \frac{dz}{2\pi i}\sum_k \frac{z^{n+1}}{(z-1)^{k+2}}
\langle
\tV_\infty,L_k\tV_1,V_0
\rangle
+\oint_0 \frac{dz}{2\pi i}\sum_k \frac{z^{n+1}}{z^{k+2}}
\langle
\tV_\infty,\tV_1,L_k\tV_0
\rangle
\nt&&\back
=
\langle
\tV_\infty,
 [L_{-1}+(n+1)L_0+
  \sum_{i=1}^n \textstyle{\frac{(n+1)!}{(i+1)!(n-i)!}}L_{i}]\tV_1,
 \tV_0
\rangle
+\langle
\tV_\infty,\tV_1,L_{n}\tV_0
\rangle\,.~~~~~~\ea
In particular, for $n=0$ case, we obtain
\ba \label{L1}
\langle
\tV_\infty,L_{-1}\tV_1,\tV_0
\rangle
=
\langle L_0\tV_\infty,\tV_1,\tV_0\rangle
-\langle \tV_\infty,L_0\tV_1,\tV_0\rangle
-\langle \tV_\infty,\tV_1,L_0\tV_0\rangle\,.
\ea
Using this relation, we can remove the $L_{-1}$ operator in eq.\,(\ref{n00}):
\ba \label{Ln}
\langle
L_{-n} \tV_\infty,\tV_1,\tV_0
\rangle
&=&
\langle L_0\tV_\infty,\tV_1,\tV_0 \rangle
+\langle \tV_\infty,\tV_1,(L_n-L_0)\tV_0 \rangle
\nt&&
+\langle
\tV_\infty,
 (nL_0+
  \sum_{i=1}^n \textstyle{\frac{(n+1)!}{(i+1)!(n-i)!}}L_{i})\tV_1,
 \tV_0
\rangle\,.
\ea
On the right hand side, there are only the operators $L_n$ with non-negative $n$.
So, use of the highest weight condition and the commutation relations
will simplify the expression further.

In similar ways, we can derive the recursion formulae for the action of $L_{-n}$ on $\tV_1$ and $\tV_0$:
\ba \label{00n}
&&\back
\langle
\tV_\infty,\tV_1,L_{-n}\tV_0
\rangle
~=~
\langle (L_n-L_0)\tV_\infty,\tV_1,\tV_0\rangle
+\langle \tV_\infty,\tV_1,L_0\tV_0\rangle
\nt&&\qqqquad\!
+\langle \tV_\infty,(nL_0+\sum_{i=1}^\infty \textstyle{
 \frac{(-1)^i(n-1+i)!}{(i+1)!(n-2)!}}L_{i})\tV_1,\tV_0\rangle
\,,
\ea\ba \label{0n0}
&&\back
\langle
\tV_\infty,L_{-n}\tV_1,\tV_0
\rangle
\nt&&\back
=
\langle
[(-1)^{n+1}(L_0-L_1)
 +L_n
 +\sum_{j=0}^\infty \tfrac{(n-1+j)!}{(j+1)!(n-2)!}L_{n+1+j}]
 \tV_\infty,\tV_1,\tV_0
\rangle
\nt
&&\back\quad
+(-1)^n\langle
\tV_\infty,\tV_1,
 (
 nL_0+\sum_{i=1}^\infty \textstyle{\frac{(n-1+i)!}{(i+1)!(n-2)!}}L_{i}) \tV_0
\rangle
+(-1)^n\langle \tV_\infty,L_0\tV_1,\tV_0\rangle
\,.\qquad\ea

\sss{Recursion formula for $W$ insertion}

We start again from the standard conformal bootstrap for $\Gamma$-type
3-point function:
\ba \label{Wn00}
&&\back
\langle
W_{-n}\tV_\infty,\tV_1,\tV_0
\rangle
\nt&&\back
=
\langle
\tV_\infty,
 [W_{-2}+(n+2)W_{-1}+\tfrac12{(n+1)(n+2)}W_{0}
  +\sum_{i=1}^n\textstyle{\frac{(n+2)!}{(i+2)!(n-i)!}}W_i]
\tV_1,\tV_0
\rangle
\nt&&
+\langle
\tV_\infty,\tV_1,W_{n}\tV_0
\rangle
\,.\ea
In particular, for  $n=0$ case, we obtain
\ba \label{W2}
\langle
\tV_\infty,W_{-2}\tV_1,\tV_0
\rangle
&=&
\langle W_0\tV_\infty,\tV_1,\tV_0\rangle
-\langle \tV_\infty,W_0\tV_1,\tV_0\rangle
-\langle \tV_\infty,\tV_1,W_0\tV_0\rangle
\nt&&
-2\langle \tV_\infty,W_{-1}\tV_1,\tV_0\rangle
\,.\ea
As in the recursion formula for $T$ insertion, we combine it with the original general formula (\ref{Wn00}), then
\ba \label{Wn}
(\ref{Wn00})
&=&
\langle W_0\tV_\infty,\tV_1,\tV_0\rangle
+\langle \tV_\infty,\tV_1,(W_n-W_0)\tV_0\rangle
\nt&&
+\langle \tV_\infty,[nW_{-1}+\tfrac12n(n+3)W_0
 +\sum_{i=1}^n \textstyle{\frac{(n+2)!}{(i+2)!(n-i)!}}W_i]
\tV_1,\tV_0\rangle
\,.~~~~\ea
This is simpler than the original one, since we don't have $W_{-2}$ insertion.
On the other hand, unlike the $T$ insertion, there remains $W_{-1}$ insertion
which makes the analysis fundamentally difficult.
In fact, as we will see later, we need impose the level-1 null state condition
on one of the vertex operators to solve the recursion.

Similarly, the formula for insertion of $W_{-n}$ in $\tV_1$ or $\tV_0$ is given as
\ba
&&\back\langle
\tV_\infty,\tV_1,W_{-n}\tV_0
\rangle
~=~
\langle (W_n-W_0)\tV_\infty,\tV_1,\tV_0\rangle
+\langle \tV_\infty,\tV_1,W_0\tV_0\rangle
\nt&&\qquad
+\langle \tV_\infty,[nW_{-1}-\tfrac12 n(n-3)W_0
 -\sum_{i=1}^\infty
  \textstyle{\frac{(-1)^i(n-1+i)!}{(i+2)!(n-3)!}}W_{i}]\tV_1,\tV_0\rangle\,,\quad
\label{W00n}
\ea\ba
&&\back
\label{W0n0}
\langle
\tV_\infty,W_{-n}\tV_1,\tV_0
\rangle
\nt&&\back
=
\langle
 [W_n
 +(n-2)W_{n+1}
 +\sum_{j=0}^\infty \tfrac{(n-1+j)!}{(j+2)!(n-3)!}W_{n+2+j}]
 \tV_\infty,\tV_1,\tV_0
\rangle
\nt&&
+(-1)^{n}\langle
 [(n-1)W_0-(n-2)W_1-W_2]\tV_\infty,\tV_1,\tV_0
\rangle
\nt&&
-(-1)^{n}\langle
\tV_\infty,\tV_1,
 [\tfrac12 n(n-1)W_0
  +\sum_{i=1}^\infty\textstyle{\frac{(n-1+i)!}{(i+2)!(n-3)!}}W_{i}] \tV_0
\rangle
\nt&&
-(-1)^{n}\langle \tV_\infty,[nW_{-1}+(n-1)W_0]\tV_1,\tV_0\rangle
\,.\ea

\sss{General procedure}
The recursion formula obtained so far will give the following algorithm to compute the 3-point correlation functions which take the general form of
\ba\label{gen3pt}
&&\back
\langle
\cL_{-Y_\infty}V_{\valp_\infty},
\cL_{-Y_1}V_{\valp_1},
\cL_{-Y_0}V_{\valp_0}
\rangle
\ea
where $V$'s are the primary fields.  It is given by repeating following steps:
\begin{enumerate}
\item Scan the leftmost operators acting on each primary field.
If you find an operator of the form $L_{n}$ ($n<0$) or $W_{n}$ ($n<-1$),
apply one of the appropriate formula in eqs.\,(\ref{L1})--(\ref{0n0}) and
eqs.\,(\ref{W2})--(\ref{W0n0}) which adds the leftmost operator
$L_n$ ($n\geq 0$) or $W_n$ ($n\geq -1$)
to different entries in the 3-point function.
\item Apply the commutation relation (\ref{comm}) to make each entry normal ordering,
and the highest weight condition (\ref{highest}) for each vertex operator.
\end{enumerate}
In each step, the degree of operators $|Y_\infty|+|Y_1|+|Y_0|$ decreases
monotonously. In the end, eq.\,(\ref{gen3pt}) becomes
the linear combination of functions of the form
\ba\label{deadE}\qqquad
\langle V_{\valp_\infty},
(W_{-1})^{\ell} V_{\valp_1},
V_{\valp_0}\rangle
\qqquad(\ell=0,1,2,\cdots)
\ea
with the coefficients depending on $\Delta$ and $w$ of the vertex operators.

For the generic $V_{\valp_\infty}$, $V_{\valp_1}$, $V_{\valp_0}$,
this is the dead end of the recursion formula.
That is, the expressions (\ref{deadE}) with $\ell=1,2,\cdots$ remain undetermined,
and the correlation function is generally written in terms of these quantities.

However,
for the computation of the correlation functions of Toda theory
associated with the linear quiver gauge theory,
we have the extra condition that one of the vertex operators,
say $V_{\valp_1}$, satisfies $\valp_1=\gamma\vlam_1$ or $-\gamma\vlam_3$.
The vertex operator of this form corresponds to the `simple' puncture
on Seiberg-Witten curve, and it is the level-1 degenerate state
which satisfies
\ba\label{null}
W_{-1}V_{\valp_1}=\frac{3w_{\valp_1}}{2\Delta_{\valp_1}}L_{-1}V_{\valp_1}\,.
\ea
Use of this condition reduces the number of $W_{-1}$ in the correlator (\ref{deadE}).
After we use the above reduction algorithm again and again,
all the $W_{-1}$ operators can be removed in the end.
Then we are left with the linear combination of
the single correlator
$\langle V_{\valp_\infty}, V_{\valp_1}, V_{\valp_0}\rangle$
whose coefficients are some functions of $\Delta$ and $w$.
The concrete form of this correlator is already given in eqs.\,(\ref{C})--(\ref{C2}).

It means that by using this algorithm, we can evaluate general 3-point functions which appear in the following discussion.

\section{Check on AGT-W relation}
\label{sec:sol}

In the previous section, we study the general procedure to
calculate $(n+3)$-point correlation functions in $A_2$ Toda theory.
Their general form is
\ba\label{gencor}
&&
\left\lag
 V_{\vbet_\infty}(\infty)
 V_{\vbet_{n+1}}(1)
 \left(
  \prod_{k=1}^n V_{\vbet_k}(q_1\cdots q_k)
 \right)
 V_{\vbet_0}(0)
\right\rag
\nt&&=
 \sum_{\{\valp_k\}}\sum_{|Y_k|=|Y'_k|}
 \left(\prod_{k=1}^n q_k^{\Delta_{\valp_k}}\right)
 \left(\prod_{k=1}^{n} (q_1\cdots q_k)^{-\Delta_{\vbet_k}}
  (q_1\cdots q_n)^{-\Delta_{\vbet_0}}\right)
\nt&&\quad\times\,
 \Gamma'_{\vbet_\infty,\vbet_{n+1},\valp_1}(\emp,\emp,Y_1)
 \left(\prod_{k=1}^{n-1}
  q_k^{|Y_k|} S^{-1}_{\valp_k}(Y_k,Y'_k)
  \Gamma_{\valp_k,\vbet_k,\valp_{k+1}}(Y'_k,\emp,Y_{k+1})
 \right)
\nt&&\quad\times\,
 q_n^{|Y_n|} S^{-1}_{\valp_n}(Y_n,Y'_n)
 \Gamma_{\valp_n,\vbet_n,\vbet_0}(Y'_n,\emp,\emp)
\ea
where $S_\valp^{-1}$ is the inverse Shapovalov matrix,
$\Gamma_{\valp_1,\valp_2,\valp_3}$
are the 3-point functions.
Here 
we rewrite the positions of punctures as
\ba
z_k=q_1\cdots q_k \quad (k=1,\cdots,n)\,,
\ea
since in the $SU(2)$ case of original AGT relation,
these $q_k$'s are identified with the coupling constants $\tau_k$
of quiver gauge theory as $q_k=e^{2\pi i\tau_k}$.

Now we calculate the 5-point correlation functions,
and check whether or not the AGT-W relation are satisfied.
According to the AGT-W conjecture,
these functions should correspond to the partition functions of
the $SU(3)\times SU(3)$ or $SU(3)\times SU(2)$ quiver gauge theory.

\subsection{$SU(3)\times SU(3)$ quiver}

In this subsection, we discuss the $SU(3)\times SU(3)$ case.
The corresponding 5-point correlation function is that of the following diagram:
\ba\label{diagram}
{\setlength{\unitlength}{1cm}
\begin{picture}(4,1.7)
 \put(0,0){\line(1,0){4}}
 \put(1,0){\line(0,1){1}}
 \put(2,0){\line(0,1){1}}
 \put(3,0){\line(0,1){1}}
 \put(-0.7,0){\small$\vbet_\infty$}
 \put(0.8,1.2){\small$\vbet_3$}
 \put(1.8,1.2){\small$\vbet_1$}
 \put(2.8,1.2){\small$\vbet_2$}
 \put(4.2,0){\small$\vbet_0$}
 \put(1.3,0.15){\small$\valp_1$}
 \put(2.3,0.15){\small$\valp_2$}
\end{picture}}
\ea
The momenta $\valp$'s and $\vbet$'s are of the form:
\ba\label{setparam}
&&
\valp_j = Q\vrho+i\valp'_j \quad(j=1,2)\,,\quad
\vbet_k = Q\vrho+i\vbet'_k \quad(k=0,\infty)\,,
\nt&&
\vbet_1 = (Q/2+im_1) (-3\vlam_3)\,,\quad 
\vbet_2 = (Q/2+im_2)\cdot 3\vlam_1\,,
\nt&&
\vbet_3 = (Q/2+im_3) (-3\vlam_3)\,,
\ea
where
$\sum_{p=1}^3 \alpha'_{j,p}=\sum_{p=1}^3 \beta'_{k,p}=0$,
and all the parameters $\valp_j$, $\vbet_k$, $m_i$ are real.
Here we naturally set the momenta
$\vbet_2\propto\vlam_1$ and $\vbet_3\propto-\vlam_3$,
since they correspond to the mass of fundamental and antifundamental fields in the gauge theory, respectively.
On the other hand, $\vbet_1$ corresponds to
the mass of bifundamental field, so we don't have
prescription for it at this point. 
It will be fixed later by requirement of AGT-W relation.
Here we set $\vbet_1\propto-\vlam_3$ in eq.\,(\ref{setparam}).
We note that it is also possible to set $\vbet_1\propto\vlam_1$ for
the following discussion without changing the result.

In the following, we calculate this 5-point correlation function,
which can be written as the linear combination of
\ba\label{VYY}
V_{(|Y_1|,|Y_2|)}
\!\!&:=&\!\!
\sum_{|Y_1|=|Y_1'|}\sum_{|Y_2|=|Y_2'|}
q_1^{|Y_1|} q_2^{|Y_2|}
\Gamma'_{\vbet_\infty,\vbet_3,\valp_1}(\emp,\emp,Y_1)
S^{-1}_{\valp_1}(Y_1,Y_1')
\nt&&\quad\times
\Gamma_{\valp_1,\vbet_1,\valp_2}(Y_1',\emp,Y_2)
S^{-1}_{\valp_2}(Y_2,Y_2')
\Gamma_{\valp_2,\vbet_2,\vbet_0}(Y_2',\emp,\emp)\,,
\ea
according to eq.\,(\ref{gencor}).
Hereafter, the descendant level of $V_{(|Y_1|,|Y_2|)}$
is denoted as $[|Y_1|,|Y_2|]$.

\sss{1-loop part}

For 1-loop part, the internal states are also primary fields.
Then the correlation function is given as
\ba\label{Vemp}
V_{(0,0)}
=
\Gamma'_{\vbet_\infty,\vbet_3,\valp_1}(\emp,\emp,\emp)\,
\Gamma_{\valp_1,\vbet_1,\valp_2}(\emp,\emp,\emp)\,
\Gamma_{\valp_2,\vbet_2,\vbet_0}(\emp,\emp,\emp)
\ea
Using the explicit form of 3-point functions (\ref{C})--(\ref{C2}) and momenta (\ref{setparam}),
and the properties of Upsilon function (\ref{G1})--(\ref{G0}),
eq.\,(\ref{Vemp}) becomes
\ba
V_{(0,0)}
\!\!\!&=&\!\!\!
f(-\vbet_\infty)f(\vbet_0)\tilde{g}(m_1)g(m_2)\tilde{g}(m_3)
\nt&&\!\!\!\times
\prod_{l=1,2}\prod_{p<q}(\alpha'_{l,p}-\alpha'_{l,q})^2
\left| \G2(i\alpha'_{l,p}-i\alpha'_{l,q}+b)\G2(i\alpha'_{l,p}-i\alpha'_{l,q}
+\frac1b) \right|^{-2}
\nt&&\!\!\!\times
\prod_{p,q=1}^3
 \biggl|\G2(\frac{Q}{2}+i\alpha'_{1,p}-i\beta'_{\infty,q}+im_3)\biggr|^2
 \biggl|\G2(\frac{Q}{2}+i\alpha'_{1,p}-i\alpha'_{2,q}-im_1) \biggr|^{2}
\nt&&\qquad\quad\times
 \biggl|\G2(\frac{Q}{2}+i\alpha'_{2,p}-i\beta'_{0,q}+im_2) \biggr|^{2}
\label{1loop}
\,,\ea
up to the factors which only depend on $b$. 
Here we define
\ba
f(\vbet)&:=&\left[\pi \mu \gamma(b^2)b^{2-2b^2} \right]^{-\vbet\cdot\vrho/b}
\prod_{e>0} \Up((Q\vrho-\valp) \cdot \vec{e}) \nt
g(m)&:=&\left[\pi \mu \gamma(b^2)b^{2-2b^2} \right]^{-3(Q/2+im)\vlam_1\cdot\vrho/b} \Up(3(Q/2+im)) \nt
\tilde{g}(m)&:=&\left[\pi \mu \gamma(b^2)b^{2-2b^2} \right]^{-3(Q/2+im)\vlam_3\cdot\vrho/b} \Up(-3(Q/2+im)) \,.
\ea

Now we show that $V_{(0,0)}$ corresponds to the 1-loop part of the partition function (\ref{Z1lp}) for $SU(3)\times SU(3)$ gauge theory,
when the parameters of the Toda theory and the gauge theory are identified appropriately.
First, note that if we identify
\ba\label{VEV}
b=\eps_1\,,\quad \frac1b=\eps_2\,,\quad
i\valp'_l=\vec{\hat{a}}_l~:~
\mbox{VEV's of adjoint scalars}\,,
\ea
the second line of eq.\,(\ref{1loop}) is equal to the product of
$\prod_{l}|z_{\rm vec}^{\rm 1lp}(\vec{\hat{a}}_l)|^2$
and the van der Monde factor.
$\prod_{l}\prod_{j<k}|\hat{a}_{l,j}-\hat{a}_{l,k}|^2$
This is the natural integral measure in the Coulomb branch.
Next, if we identify
\ba\label{mass}
\bar\mu_\bp=\frac{Q}{2}+im_3-i\beta'_{\infty,\bp}\,,\quad
m=\frac{Q}{2}+im_1\,,\quad
\mu_p=\frac{Q}{2}-im_2+i\beta'_{0,p}\,,
\ea
for the mass of antifundamental, bifundamental and fundamental fields,
the factors in the third and fourth lines of (\ref{1loop}) are equal to
$\prod_{\bp}|z_{\rm afd}^{\rm 1lp}(\vec{a}_1,\bar{\mu}_\bp)|^2$,
$|z_{\rm bfd}^{\rm 1lp}(\vec{a}_1,\vec{a}_2,m)|^2$ and
$\prod_{p}|z_{\rm fd}^{\rm 1lp}(\vec{a}_2,\mu_p)|^2$, respectively.

To summarize,
with the identification of parameters (\ref{VEV})--(\ref{mass}),
the correlation function at this level can be written as,
up to some factors,
\ba
V_{(0,0)}=\left|Z_\tx{1-loop}\right|^2
\ea
where $Z_\tx{1-loop}$
is the 1-loop factor of Nekrasov partition function for $SU(3)\times SU(3)$ quiver gauge theory.
This is the AGT-W relation for the 1-loop part.

\sss{Instanton part}

For instanton part, we check the AGT-W relation for each instanton
level $[|Y_1|,|Y_2|]$ in eq.\,(\ref{VYY}).

\sssudl{Level $[n,0]$ or $[0,n]$}

For these levels, the discussion is almost parallel to
4-point function's case, which corresponds to AGT-W relation for $SU(3)$
gauge theory~\cite{Mironov:2009by},
since one of the intermediate operators is set to be primary.
For simple examples, the partition functions for $SU(3)_1\times SU(3)_2$
quiver gauge theory at the instanton level [1,0] and [0,1] are
\ba
 \label{inst10}
Z_\text{inst}^{([1],\emp)}
\!\!&=&\!\!
q_1\sum_{i=1}^3\frac
 {\prod_{\bp=1}^3(\ha_{1,i}+\bmu_\bp)\prod_{j=1}^3(\ha_{1,i}-\ha_{2,j}-m)}
 {\prod_{i\neq k}(\ha_{1,i}-\ha_{1,k})(\ha_{1,i}-\ha_{1,k}+\eps_+)}
\nt\!\!&=:&\!\!
\sum_{i=1}^3 R_{1,i}(\ha_{1,i})\,,\\
 \label{inst01}
Z_\text{inst}^{(\emp,[1])}
\!\!&=&\!\!
q_2\sum_{i=1}^3\frac
 {\prod_{j=1}^3(\ha_{1,j}-\ha_{2,i}-m+\eps_+)\prod_{p=1}^3(\ha_{2,i}-\mu_p+\eps_+)}
 {\prod_{i\neq k}(\ha_{2,i}-\ha_{2,k})(\ha_{2,i}-\ha_{2,k}+\eps_+)}
\nt\!\!&=:&\!\!
\sum_{i=1}^3 R_{2,i}(\ha_{2,i})\,,
\ea
where $\eps_+=\eps_1+\eps_2$ is identified with $Q$ from eq.\,(\ref{VEV}).
Therefore, eq.\,(\ref{inst10}) is the same form as the level-1 instanton partition
function for $SU(3)_1$ gauge theory with the mass of fundamental fields $\mu_j=Q/2-im_1+i\hat\alp_{2,j}$.
Similarly, eq.\,(\ref{inst01}) corresponds to that for $SU(3)_2$ theory
with the mass of antifundamental fields
$\bmu_j=Q/2+im_1-i\hat\alp_{1,j}$.
Compared with eq.\,(\ref{mass}), they are exactly the relevant 4-point functions
which is obtained
by cutting one of internal lines $\valp_2/\valp_1$
in our 5-point function (\ref{diagram})
and regarding it as an external line.
It has been already shown that these 4-point functions correspond to some conformal
blocks of $A_2$ Toda theory~\cite{Mironov:2009by}.

Now we also have to mention the $U(1)$-factors~\cite{Alday:2009aq}.
In the setup of AGT-W relation, strictly speaking,
we discuss $U(3)\times U(3)$ quiver gauge theory.
The $U(1)$ part of gauge symmetry is actually decoupled,
but the $U(1)$ flavor symmetry remains, which causes additional $U(1)$-factors.
Therefore,
unless we multiply it by Nekrasov partition function for $SU(3)\times SU(3)$,
we cannot establish the AGT-W relation.
In this case, the $U(1)$-factor is of the form
\ba
Z_{U(1)}=(1-q_1)^{\nu_1}(1-q_2)^{\nu_2}(1-q_1q_2)^{\nu_3},
\ea
where
\ba\label{u1}
&&\back
\nu_1=-3\left(\frac{Q}{2}+im_3\right)\left(\frac{Q}{2}-im_1\right)
,\quad
\nu_2=-3\left(\frac{Q}{2}+im_1\right)\left(\frac{Q}{2}+im_2\right)
,\nt
&&\back
\nu_3=-3\left(\frac{Q}{2}+im_3\right)\left(\frac{Q}{2}+im_2\right)
.\ea

On the other hand, the correlation function of $A_2$ Toda theory
at these levels are
\ba
V_{(1,0)}
&=&
\sum_{Y_1,Y_1'}
\frac{
 \Gamma'_{\vbet_\infty,\vbet_3,\valp_1}(\emp,\emp,Y_1)
 S^{-1}_{\valp_1}(Y_1,Y_1')
 \Gamma_{\valp_1,\vbet_1,\valp_2}(Y_1',\emp,\emp)
}{
 \Gamma_{\vbet_\infty,\vbet_3,\valp_1}(\emp,\emp,\emp)
 \Gamma_{\valp_1,\vbet_1,\valp_2}(\emp,\emp,\emp)}
\,V_{(0,0)}\,,
\nt
V_{(0,1)}
&=&
\sum_{Y_2,Y_2'}
\frac{
 \Gamma_{\valp_1,\vbet_1,\valp_2}(\emp,\emp,Y_2)
 S^{-1}_{\valp_2}(Y_2,Y_2')
 \Gamma_{\valp_2,\vbet_2,\vbet_0}(Y_2',\emp,\emp)
}{
 \Gamma_{\valp_1,\vbet_1,\valp_2}(\emp,\emp,\emp)
 \Gamma_{\valp_2,\vbet_2,\vbet_0}(\emp,\emp,\emp)}
\,V_{(0,0)}\,,
\ea
where $Y_1,Y_1',Y_2,Y_2'\in\{([1],\emp),(\emp,[1])\}$.
After everything is put together, we obtain the expected result,
that is,
\ba\label{01-10}
\frac{V_{(1,0)}}{V_{(0,0)}}
=Z_\text{inst}^{([1],\emp)}+\nu_1\,,
\quad
\frac{V_{(0,1)}}{V_{(0,0)}}
=Z_\text{inst}^{(\emp,[1])}+\nu_2\,.
\ea
It is straightforward to discuss the similar relations
at level $[n,0]$ or $[0,n]$ $(n>1)$ by computer.
In fact, we have checked them for $n=2, 3$.

\sssudl{Level $[n_1, n_2]$ with $n_1, n_2>0$}

From above discussion, we identified all the parameters
of 4-dim $SU(3)$ quiver gauge theory and 2-dim $A_2$ Toda theory,
and obtained all necessary relations between them,
{\em i.e.}~eqs.\,(\ref{VEV}), (\ref{mass}) and (\ref{u1}),
for checking AGT-W relation.
With these identification of parameters,
we  carry out the check of the relation for
 the instanton level $[n_1,n_2]$ $(n_1,n_2>0)$.

The simplest example is level [1,1].
The partition function and $U(1)$ factor of gauge theory
at this level are
\ba
Z_\tx{inst}^{([1],[1])}
\!\!&=&\!\!
\sum_{i,j=1}^3
R_{1,i}(\ha_{1,i})\,R_{2,j}(\ha_{2,j})
\frac{(\ha_{1,i}-\ha_{2,j}-m+\eps_1)(\ha_{1,i}-\ha_{2,j}-m+\eps_2)}
 {(\ha_{1,i}-\ha_{2,j}-m+\eps_+)(\ha_{1,i}-\ha_{2,j}-m)}\,,\nt
Z_{U(1)}^{[1,1]}
\!\!&=&\!\!
 \nu_1 Z_\tx{inst}^{(0,[1])}+\nu_2 Z_\tx{inst}^{([1],0)}
 +\nu_1\nu_2+\nu_3\,.
\ea
On the other hand, the correlation function of Toda theory at this level is
\ba
V_{(1,1)}\!\!\!&=&\!\!\!
\sum_{Y_1,Y_1',Y_2,Y_2'}
 \Gamma_{\vbet_\infty,\vbet_3,\valp_1}(\emp,\emp,Y_1)
 S^{-1}_{\valp_1}(Y_1,Y_1')
 \Gamma_{\valp_1,\vbet_1,\valp_2}(Y_1',\emp,Y_2)
\nt&&\qquad\quad\,\times\,
 S^{-1}_{\valp_2}(Y_2,Y_2')
 \Gamma_{\valp_2,\vbet_2,\vbet_0}(Y_2',\emp,\emp)\,,
\ea
where $Y_1,Y_1',Y_2,Y_2'\in\{([1],\emp),(\emp,[1])\}$.
Then we can obtain the expected result, that is,
\ba
\frac{V_{(1,1)}}{V_{(0,0)}}
=Z_\tx{inst}^{([1],[1])}+
Z_{U(1)}^{[1,1]}\,.
\ea
We have also successfully checked the $(n_1,n_2)=(1,2),(2,1)$ cases
by computer. Since the formula become complicated, we don't write
their explicit form here.

\sss{Summary}

We check the AGT-W relation
in $SU(3)\times SU(3)$ quiver case for the 1-loop part
and the instanton corrections for the level $[|Y_1|, |Y_2|]$ with $|Y_1|+|Y_2|\leq 3$.
While this is only partial result toward the proof, the coincidence of
the explicit formula in both side is quite nontrivial and very convincing.
Its generalization to the case of
$SU(3)\times\cdots\times SU(3)$ quiver seems straightforward.
Moreover, for 1-loop part, it is easy to generalize to the case of
$SU(N)\times\cdots\times SU(N)$ quiver, where the argument is a
straightforward generalization of the $SU(3)$ quiver case.

\subsection{$SU(3)\times SU(2)$ quiver}
\label{SU3SU2}

As we mentioned before,
the other kind of 5-point correlation function of $A_2$ Toda theory
should correspond to the partition function of $SU(3)\times SU(2)$
quiver gauge theory. Let us now discuss the AGT-W relation in this case.

\sss{1-loop part}

The puncture $V_{\vbet_0}(0)$ now becomes the `simple' puncture,
instead of `full' puncture.
So we must consider the correlation function of the diagram (\ref{diagram}) with
$\vbet_0=(Q/2+im_0)\cdot 3\vlam_1$.

After having set this value, we meet an immediate problem.
To see this, the last factor in eq.\,(\ref{VYY}) contains two
vertices $\beta_0$, $\beta_2$ to be proportional to $\vec\lambda_1$.
This $\Gamma$ factor, at the level $|Y_2'|=0$, is written by the formula (\ref{C1})
where one of the vector, say $\vec\alpha_2$, is proportional to $\vec\lambda_1$.\footnote{
Here, please don't confuse this $\valp_2$ in eq.\,(\ref{C1}) with that in eq.\,(\ref{diagram}).}
Then in the numerator, we have a factor
$\Upsilon((Q\vec\rho-\vec\alpha_2)\cdot \vec e_2)=\Upsilon(Q)$,
which vanishes because of the property of $\Upsilon$ function.
Such zero factors always exist for the linear quiver
whose product gauge groups have different ranks.
While it may imply a limitation to AGT-W conjecture,
in the following study, we drop such factors since they don't
depend on the momentum of the intermediate operator.
In appendix \ref{s:3pt}, we derive some properties of 3-point functions
where two of the three operators have level-1 singular state,
especially the constraint for the third operator to have non-vanishing
3-point function. 
As it is explained there, from the conformal
Ward identities, we need a relation between $\Delta$ and $w$ for the
third generator.  This is, however, a much weaker condition than
eqs.\,(\ref{C1})--(\ref{C2}), where the correlation function vanishes for arbitrary
vertex.  This may imply that we need some modifications in such special cases.
In this paper, however, we will not try to go  further in this direction,
but simply drop the zero factor and study the consequence.
This may be justified, since it does not depend on the parameters of the theory.

For the internal momentum $\valp_2$ in eq.\,(\ref{diagram}),
we must set it 
as
a 1-parameter vector, since it will correspond to the VEV of $SU(2)$ gauge group.
There is arbitrariness to choose the form of $\valp_2$, but a natural choice will be
$\valp_2=Q\vrho+i(\alpha'_2,-\alpha'_2,0)$ or $Q\vrho+i(\alpha'_2,\alpha'_2,-2\alpha'_2)$.
If we adopt the later one, van der Monde factor becomes $0$.
This is not desirable, so we use the former one.
We expect that $V_{(0,0)}$ corresponds to the 1-loop part of the partition function
for $SU(3)\times SU(2)$ quiver gauge theory. In order to achieve this, however, we
are forced  to impose an additional condition $m_2+m_0=0$.
It helps to cancel out unnecessary factors of $V_{(0,0)}$ and we obtain
\ba
V_{(0,0)}&=&\prod_{p<q}^3(\alpha'_{1,p}-\alpha'_{1,q})^2
\left| \G2(i\alpha'_{1,p}-i\alpha'_{1,q}+b)\G2(i\alpha'_{1,p}-i\alpha'_{1,q}
+\frac1b) \right|^{-2} \nt
&&\times 4\alpha'^2_{2} \left| \G2(2i\alpha'_2+b)\G2(2i\alpha'_2+\frac1b) \right|^{-2} \nt
&&\times \prod_{p,q=1}^3
\biggl|\G2(\frac{Q}{2}+i\alpha'_{1,p}-i\beta'_{\infty,q}+im_3)\biggr|^2
\prod_{p=1}^3 \biggl|\G2(\frac{Q}{2}+i\alpha'_{1,p}+im_1)\biggr|^2 \nt
&&\times\prod_{p=1}^3 \prod_{q=1}^2\biggl|\G2(\frac{Q}{2}+
i\alpha'_{1,p}-i\alpha'_{2,q}-im_1) \biggr|^{2}\cdot\nt
&&~~~\cdot\biggl|\G2(\frac{Q}{2}-i\alpha'_2-3im_2)\G2(\frac{Q}{2}+i\alpha'_2-3im_2) \biggr|^{2}
\label{3-2}
\ea
up to some constant which is independent of $\valp_1$ and $\valp_2$ .
As in the previous subsection, we can see that if we identify
\ba
i\valp'_1&=&\vec{\hat{a}}_1~:~\mbox{VEV's of $SU(3)$ adjoint scalar} \nt
(i\alpha'_2,-i\alpha'_2)&=&\vec{\hat{a}}_2~:~\mbox{VEV's of $SU(2)$ adjoint scalar},
\ea
the first line of eq.\,(\ref{3-2}) is equal to the product of
$|z_{\rm vec}^{\rm 1lp}(\vec{\hat{a}}_1)|^2$
and van der Monde factor
$\prod_{j<k}|\hat{a}_{1,j}-\hat{a}_{1,k}|^2$ for $SU(3)$ gauge group
and the second line is equal to that for $SU(2)$ gauge group.
Also, we can find that if we identify
\ba
\bar\mu_\bp=\frac{Q}{2}+im_3-i\beta'_{\infty,\bp}\,,\quad
\nu=\frac{Q}{2}+im_1\,,\quad
m=\frac{Q}{2}+im_1\,,\quad
\mu=\frac{Q}{2}+3im_2\,,
\ea
with the mass of three $SU(3)$ antifundamental, a $SU(3)$ fundamental, a $SU(3) \times SU(2)$ bifundamental
and a $SU(2)$ fundamental fields,
the factors in the third and  fourth lines of eq.\,(\ref{3-2}) equal to
$\prod_{\bp}|z_{\rm afd}^{\rm 1lp}(\vec{a}_1,\bar{\mu}_\bp)|^2$,
$|z_{\rm fd}^{\rm 1lp}(\vec{a}_1,\nu)|^2$,
$|z_{\rm bfd}^{\rm 1lp}(\vec{a}_1,\vec{a}_2,m)|^2$ and
$|z_{\rm fd}^{\rm 1lp}(\vec{a}_2,\mu)|^2$, respectively.

Therefore, we see that eq.\,(\ref{3-2}) can be written as $V_{(0,0)}=\left|Z_\tx{1-loop}\right|^2$,
where $Z_\tx{1-loop}$ is the 1-loop factor of Nekrasov partition function
for $SU(3)\times SU(2)$ quiver gauge theory, up to the zero factors which we mentioned above.
Here we have to note, however,
we have met additional problems in the case of $SU(3) \times SU(2)$ quiver.
First, $SU(3)$ fundamental field and $SU(3) \times SU(2)$
bifundamental field have the same mass $\nu=m=Q/2+im_1$.
In the gauge theory side, it is not necessary for the two fields to have the same mass.
How can we make these two mass independent in Toda theory?
Secondly, we need to impose the condition $m_2+m_0=0$ to get the correspondence.
This is artificial and there seems no physical meaning in Toda theory.
We don't know a correct answer yet, but one possibility to resolve this puzzle
may be that the form of the 3-point functions (\ref{C1})--(\ref{C2}) could be modified,
when more than one of momenta $\valp_1$ or/and  $\valp_2$
are also proportional to $\vlam_1$ or $\vlam_3$.
The representation of W-algebra for a degenerate field is very different
from that for a non-degenerate one.
Therefore, it may be possible that the form of the 3-point functions
changes when two vertex operators are
degenerate fields, which may also resolve the problem of zero factor.

\sss{Instanton part}

For the instanton level $[n,0]$, we can discuss in a similar way
to the $SU(3)\times SU(3)$ case.
In the simplest $n=1$ case, the partition function is
\ba
 \label{inst10-32}
Z_\text{inst}^{([1],\emp)}
=
q_1\sum_{i=1}^3\frac
 {\prod_{\bp=1}^3(\ha_{1,i}+\bmu_\bp)\prod_{j=1}^2(\ha_{1,i}-\ha_{2,j}-m)\cdot
  (\ha_{1,i}-\nu+\eps_+)}
 {\prod_{i\neq k}(\ha_{1,i}-\ha_{1,k})(\ha_{1,i}-\ha_{1,k}+\eps_+)}\,,
\ea
which is the same form as the level-1 instanton partition function
for $SU(3)$ gauge theory with the mass of fundamental fields
$\nu$ and $\mu_j=Q/2-im_1+i\hat\alpha_{2,j}$.
Therefore, together with $U(1)$-factor (\ref{u1}), we successfully obtain
\ba
\frac{V_{(1,0)}}{V_{(0,0)}}
=Z_\text{inst}^{([1],\emp)}+\nu_1\,.
\ea
We also have checked the correspondence in the $n=2$ case by computer.

On the other hand, however, for the level $[0,n]$, the correspondence is rather nontrivial. The partition function in the $n=1$ case is
\ba
 \label{inst01-32}
Z_\text{inst}^{(\emp,[1])}
=
q_2\sum_{i=1}^2\frac
 {\prod_{j=1}^3(\ha_{1,j}-\ha_{2,i}-m+\eps_+)\cdot(\ha_{2,i}-\mu+\eps_+)}
 {\prod_{i\neq k}(\ha_{2,i}-\ha_{2,k})(\ha_{2,i}-\ha_{2,k}+\eps_+)}\,,
\ea
which is the same form as the level-1 instanton partition function
for $SU(2)$ gauge theory.
Therefore, in order to match this function with Toda correlation function
\ba
V_{(0,1)}
=
\sum_{Y_2,Y_2'}
\frac{
 \Gamma_{\valp_1,\vbet_1,\valp_2}(\emp,\emp,Y_2)
 S^{-1}_{\valp_2}(Y_2,Y_2')
 \Gamma_{\valp_2,\vbet_2,\vbet_0}(Y_2',\emp,\emp)
}{
 \Gamma_{\valp_1,\vbet_1,\valp_2}(\emp,\emp,\emp)
 \Gamma_{\valp_2,\vbet_2,\vbet_0}(\emp,\emp,\emp)}
\,V_{(0,0)}\,,
\ea
we need to sum up only the $L_{-n}$ descendants of $V_{\valp_2}$,
{\em i.e.}~$Y_2,Y_2'\in ([1],\emp)$.
This is very strange situation,
since it means that $W_{-n}$ generators cannot live in some region of  Seiberg-Witten curve.
This region is identified as the region which a D6-brane passes through
in the intersecting D4/NS5-branes' system,
when the present $SU(3)\times SU(2)$ system is realized from the $SU(3)\times SU(3)$ system by moving a D6-brane from infinite distance.
Then if we ignore the $W_{-n}$ descendants here, we can obtain the
desirable result
\ba
\frac{V_{(0,1)}}{V_{(0,0)}}
=Z_\text{inst}^{(\emp,[1])}+\nu_2\,.
\ea
We also have checked the $n=2$ case by computer.
However, at this moment,
we have no persuasive reason to justify this procedure in Toda theory.

Finally, we mention the case of the level $[n_1,n_2]$ with $n_1,n_2>0$.
In this case, we also need to ignore the $W_{-n}$ descendants of $V_{\valp_2}$.
We have checked it in the $[1,1]$ case.

\sss{Summary}

We manage to check the AGT-W relation for $SU(3) \times SU(2)$ quiver gauge theory in the 1-loop part and the instanton part at the level $[|Y_1|,|Y_2|]$ with $|Y_1|+|Y_2|\leq 2$.
However, the correspondence is quite nontrivial,
since there remain the mysteries of the vanishing factor $\Upsilon(Q)$, the strange conditions on mass and the ignorance of $W_{-n}$ descendants.
The investigation of these problems must be an important future work.

\section{Conclusion}
In this paper, we study and confirm the AGT-W relation for
$SU(3)\times SU(3)$ gauge theory in the 1-loop
factor and also the lower level factors
in the instanton part by solving conformal Ward identity
by computer. Our calculation method is valid for the general
$SU(3)\times \cdots \times SU(3)$ cases,
and the generalization to these cases is straightforward.
Moreover, such relation seems to hold for the linear
$SU(N)\times\cdots\times SU(N)$ quiver gauge theories
as well, at least for the 1-loop part.

For the $SU(3)\times SU(2)$ quiver, however, we encountered some
problems which include the undesirable vanishing factor, the extra
conditions on the mass for bifundamental and fundamental matter fields,
and we are forced to ignore $W_{-n}$ descendants in some part of the correlation function.
For these cases, we need to have some modification to
AGT-W relation and/or
our calculation method by the decomposing Gaiotto curve.
After these problems are solved, we will be able to straightforwardly
confirm the general $SU(3)\times \cdots \times SU(3)\times SU(2)$ and
$SU(2)\times SU(3)\times \cdots \times SU(3) \times SU(2)$ case,
that is, all the case of AGT-W relation for linear $SU(3)$ quivers.
Therefore, we hope to clarify them in the future paper.

\subsection*{Acknowledgments}
We would like to thank Yuji Tachikawa for his collaboration at the early stage.
Y. M. is partially supported by KAKENHI (\#20540253) from MEXT, Japan.
S. S. is partially supported by Grant-in-Aid for Scientific Research (B) \#19340066 from MEXT, Japan.

\appendix

\section{Properties of double Gamma function $\Gamma_2$ and Upsilon function $\Upsilon$}
\label{sec:app}

The double Gamma function $\Gamma_2(x|\eps_1,\eps_2)$ is defined as
\ba
\Gamma_2(x|\eps_1,\eps_2)
=\exp\frac{d}{ds}\biggr|_{s=0}
 \zeta_2(s;x|\eps_1,\eps_2)\,,
\ea
where
\ba
\zeta_2(s;x|\eps_1,\eps_2)
:=
\sum_{m,n}(m\eps_1+n\eps_2+x)^{-s}
=
\frac{1}{\Gamma(s)}
 \int_0^\infty dt\,
 \frac{t^{s-1}e^{-tx}}{(1-e^{-\eps_1t})(1-e^{\eps_2t})}\,,
\ea
where $\Gamma(s)$ is ordinary Gamma function.
This function $\Gamma_2(x|\eps_1,\eps_2)$ is often written
as $\Gamma_2(x)$ if there is no confusion, and satisfies
the following relations:
\ba
\G2(x)^* \!\!\!&=&\!\!\! \G2(x^*) \label{G1}\\
\G2(x+\eps_1)\G2(x+\eps_2) \!\!\!&=&\!\!\!
 x\G2(x)\G2(x+\eps_1+\eps_2)\,. \label{G2}
\ea
The Upsilon function $\Upsilon(x)$ can be written as
the product of double Gamma functions:
\ba
\Up(x)=\frac{1}{\G2(x) \G2(Q-x)}\,, \label{G0}
\ea
which is also very important for AGT and AGT-W relation.

\section{Shapovalov matrix}
\label{app:Shap}

As we mentioned in the main text,
Shapovalov matrix is given as eq.\,(\ref{Shap}).
The concrete forms of its elements up to level-2 descendants
are given in eq.\,(65) of \cite{Mironov:2009by}, for example.
In this paper, we also use the concrete forms of level-3 descendants,
so we show them in this appendix.
In the following, we denote these elements (\ref{Shap}) as
$S(\cL_{-Y_1},\cL_{-Y_2})$.

{\footnotesize
\ban&&\back
S(L_{-3},L_{-3})=6\Delta+2c
\,,\quad
S(L_{-3},L_{-2}L_{-1})=10\Delta
\,,\quad
S(L_{-3},L_{-1}^3)=24\Delta
\,,\nt&&\back
S(L_{-3},L_{-2}W_{-1})=15w
\,,\quad
S(L_{-3},L_{-1}^2W_{-1})=24w
\,,\quad
S(L_{-3},L_{-1}W_{-2})=36w
\,,\nt&&\back
S(L_{-3},L_{-1}W_{-1}^2)=90D\Delta
\,,\quad
S(L_{-3},W_{-3})=9w
\,,\quad
S(L_{-3},W_{-2}W_{-1})=36D\Delta
\,,\nt&&\back
S(L_{-3},W_{-1}^3)=\tfrac{189}{2}w(3\Delta+1)
\,,\ean
\ban&&\back
S(L_{-2}L_{-1},L_{-2}L_{-1})=\Delta(8(\Delta+1)+c)
\,,\quad
S(L_{-2}L_{-1},L_{-1}^3)=12\Delta(3\Delta+1)
\,,\nt&&\back
S(L_{-2}L_{-1},L_{-2}W_{-1})=\tfrac32 w (8(\Delta+1)+c)
\,,\quad
S(L_{-2}L_{-1},L_{-1}^2W_{-1})=18w(3\Delta+1)
\,,\nt&&\back
S(L_{-2}L_{-1},L_{-1}W_{-2})=12w(\Delta+3)
\,,\quad
S(L_{-2}L_{-1},L_{-1}W_{-1}^2)=9(D\Delta(5\Delta+9)+6w^2)
\,,\nt&&\back
S(L_{-2}L_{-1},W_{-3})=21w
\,,\quad
S(L_{-2}L_{-1},W_{-2}W_{-1})=18(3D\Delta+w^2)
\,,\nt&&\back
S(L_{-2}L_{-1},W_{-1}^3)=
 \tfrac{135w}{2}
 \left(D(3\Delta+1)+\kappa(5\Delta+1)\right)
\,,\ean
\ban&&\back
S(L_{-1}^3,L_{-1}^3)=24\Delta(\Delta+1)(2\Delta+1)
\,,\quad
S(L_{-1}^3,L_{-2}W_{-1})=18w(3\Delta+1)
\,,\nt&&\back
S(L_{-1}^3,L_{-1}^2W_{-1})=36w(\Delta+1)(2\Delta+1)
\,,\quad
S(L_{-1}^3,L_{-1}W_{-2})=72w(\Delta+1)
\,,\nt&&\back
S(L_{-1}^3,L_{-1}W_{-1}^2)=54(\Delta+1)(3D\Delta+2w^2)
\,,\quad
S(L_{-1}^3,W_{-3})=60w
\,,\nt&&\back
S(L_{-1}^3,W_{-2}W_{-1})=108(D\Delta+w^2)
\,,\quad
S(L_{-1}^3,W_{-1}^3)= 81w (3D(3\Delta+4)+2w^2+1)
\,,\ean
\ban&&\back
S(L_{-2}W_{-1},L_{-2}W_{-1})=\tfrac94 D \Delta (8(\Delta+1)+c)
\,,\quad
S(L_{-2}W_{-1},L_{-1}^2W_{-1})=27(D\Delta(\Delta+1)+2w^2)
\,,\nt&&\back
S(L_{-2}W_{-1},L_{-1}W_{-2})=18(3D\Delta+w^2)
\,,\quad
S(L_{-2}W_{-1},L_{-1}W_{-1}^2)=\tfrac{27}2 D w (11\Delta+9)
\,,\nt&&\back
S(L_{-2}W_{-1},W_{-3})=\tfrac{63}2 D\Delta
\,,\quad
S(L_{-2}W_{-1},W_{-2}W_{-1})=27Dw(\Delta+3)
\,,\nt&&\back
S(L_{-2}W_{-1},W_{-1}^3)=
 \tfrac{405}{4}
 [D^2\Delta(3\Delta+1)+\kappa(D\Delta(\Delta+1)+4w^2)]
\,,\ean
\ban&&\back
S(L_{-1}^2W_{-1},L_{-1}^2W_{-1})=
 18(\Delta+1)(D\Delta(2\Delta+3)+4w^2)
\,,\nt&&\back
S(L_{-1}^2W_{-1},L_{-1}W_{-2})=
 36(D\Delta(2\Delta+3)+w^2)
\,,\nt&&\back
S(L_{-1}^2W_{-1},L_{-1}W_{-1}^2)=
 27w(D(\Delta+3)(4\Delta+3)+2w^2)
\,,\nt&&\back
S(L_{-1}^2W_{-1},W_{-3})=
 90D\Delta
\,,\quad
S(L_{-1}^2W_{-1},W_{-2}W_{-1})=
 162wD(\Delta+1)
\,,\nt&&\back
S(L_{-1}^2W_{-1},W_{-1}^3)=
 \tfrac{243}{2}
  [D^2\Delta(3\Delta+7)
  +2D(\Delta+3)w^2
  +\kappa(D\Delta(\Delta+1)+4w^2)]
\,,\ean
\ban&&\back
S(L_{-1}W_{-2},L_{-1}W_{-2})=
 \tfrac{9\kappa}{4}\Delta(8\Delta^2+(54+c)\Delta+14+c)
\,,\nt&&\back
S(L_{-1}W_{-2},L_{-1}W_{-1}^2)=
 \tfrac{27\kappa}{16}(112\Delta^2+(210-c)\Delta+34-c)
\,,\quad
S(L_{-1}W_{-2},W_{-3})=
 45\Delta(D+1)
\,,\nt&&\back
S(L_{-1}W_{-2},W_{-2}W_{-1})=
 \tfrac{27\kappa}8 w(8\Delta^2+(54+c)\Delta+14+c)
\,,\nt&&\back
S(L_{-1}W_{-2},W_{-1}^3)=
 \tfrac{243\kappa}{32}
 [w^2(48\Delta+46+c)+D\Delta(128\Delta+62+c)]
\,,\ean
\ban&&\back
S(L_{-1}W_{-1}^2,L_{-1}W_{-1}^2)=
 \tfrac{81}{2}
 [D^2\Delta(2\Delta^2+5\Delta+11)
 +\kappa D\Delta(\Delta+1)(\Delta+2)
 +4w^2(D(\Delta+3)+\kappa(\Delta+2))]
\,,\nt&&\back
S(L_{-1}W_{-1}^2,W_{-3})=
 \tfrac{135}{2}w(3D+1)
\,,\nt&&\back
S(L_{-1}W_{-1}^2,W_{-2}W_{-1})=
 81D\Delta(D(2\Delta+1)+\kappa(\Delta+1))
 +\tfrac{81\kappa}{32}w^2(48\Delta+142+c)
\,,\nt&&\back
S(L_{-1}W_{-1}^2,W_{-1}^3)=
 \tfrac{729}{4}w
 [D^2(\Delta+1)(2\Delta+5)+\kappa(D(\Delta^2+18\Delta+5)+4w^2)]
\,,\ean
\ban&&\back
S(W_{-3},W_{-3})=
 \tfrac{27}{2}D\Delta+\tfrac32(24\Delta+c)
\,,\quad
S(W_{-3},W_{-2}W_{-1})=
 \tfrac{27\kappa}{2}w(5\Delta+1)+54w
\,,\nt&&\back
S(W_{-3},W_{-1}^3)=
 \tfrac{81\kappa}{64}
 [128w^2+D\Delta(320\Delta+302+25c)]
\,,\ean
\ban&&\back
S(W_{-2}W_{-1},W_{-2}W_{-1})=
 \tfrac{81\kappa}{16}
 (32w^2+D\Delta(\Delta+1)(8\Delta+14+c))
\,,\nt&&\back
S(W_{-2}W_{-1},W_{-1}^3)=
 \tfrac{729\kappa}{64}Dw(48\Delta^2+(174+c)\Delta+62+c)
\,,\ean
\ban
S(W_{-1}^3,W_{-1}^3)\!\!\!&=&\!\!\!
 \tfrac{729}{8}
 [3D^3\Delta(\Delta+1)(2\Delta+1)
  +\kappa D^2\Delta(9\Delta^2+69\Delta+44)
  +\kappa^2D\Delta(\Delta+1)(3\Delta+5)]
 \nt\!\!\!&&\!\!\!
 +\tfrac{729\kappa}{8}w^2
 [4D(-5D+\kappa(5\Delta+1))(9\Delta+16)
 +8(4\Delta+3)]
\,,\ean
where
$D:=\kappa(\Delta+\frac15)-\frac15$,
$\kappa:=\frac{32}{22+5c}$.
}
\section{3-point functions}

In \S\ref{sec:3pt}, we show the general procedure to calculate
3-point functions (\ref{3pt}), which we must obtain in order to
check the AGT-W relation.
In this appendix, we show the concrete forms of
$\Gamma(\cL_{-Y_\infty},\cL_{-Y_0}):=
 \Gamma_{\valp_\infty,\valp_1,\valp_0}(Y_\infty,\emp,Y_0)$
at the descendant level $[|Y_\infty|,|Y_0|]$ with $|Y_\infty|+|Y_0|\leq 3$.
Those of level [1,0], [0,1], [2,0], [0,2] are already given in \cite{Mironov:2009by}.

\vspace{4mm}

{\footnotesize
\noindent\udl{Level [1,0] and [0,1]}
\ban&&\back
\Gam(L_{-1},\emp)=\Dinf+\Di-\Do
\,,\quad
\Gam(W_{-1},\emp)=\winf-\wi-\wo+\chi\Gam(L_{-1},\emp)
\,,\nt&&\back
\Gam(\emp,L_{-1})=-\Dinf+\Di+\Do
\,,\quad
\Gam(\emp,W_{-1})=-\winf+\wi+\wo-\chi\Gam(\emp,L_{-1})
\,,\ean
where $\chi:={3\wi}/{2\Di}$.

\vspace{2mm}

\noindent\udl{Level [2,0]}
\ban&&\back
\Gam(L_{-2},\emp)=\Gam(L_{-1},\emp)+\Di
\,,\quad
\Gam(L_{-1}^2,\emp)=
 \big[\Gam(L_{-1},\emp)+1\big] \Gam(L_{-1},\emp)
\,,\nt&&\back
\Gam(L_{-1}W_{-1},\emp)=
 \big[\Gam(L_{-1},\emp)+1\big] \Gam(W_{-1},\emp)
\,,\quad
\Gam(W_{-2},\emp)=
 \Gam(W_{-1},\emp)+\chi\Gam(L_{-1},\emp)
\,,\nt&&\back
\Gam(W_{-1}^2,\emp)=
 \big[\Gam(W_{-1},\emp)+\chi\big]\Gam(W_{-1},\emp)
 +\tfrac92 D_\infty\Gam(L_{-1},\emp)
\,,\ean
where
$D_\infty:=\kappa(\Dinf+\tfrac15)-\tfrac15$.

\vspace{2mm}

\noindent\udl{Level [1,1]}
\ban&&\back
\Gam(L_{-1},L_{-1})=
\Gam(L_{-1},\emp)\Gam(\emp,L_{-1})+(\Dinf-\Di+\Do)
\,,\nt&&\back
\Gam(L_{-1},W_{-1})=
\big[\Gam(L_{-1},\emp)-1\big]\Gam(\emp,W_{-1})+3\wo
\,,\nt&&\back
\Gam(W_{-1},L_{-1})=
 \Gam(W_{-1},\emp)\big[\Gam(\emp,L_{-1})-1\big]+3\winf
\,,\nt&&\back
\Gam(W_{-1},W_{-1})=
 \big[\Gam(W_{-1},\emp)-\chi\big]\Gam(\emp,W_{-1})
 +\tfrac92 D_0(\Dinf-\Di)
\,,\ean
where $D_0:=\kappa(\Do+\tfrac15)-\tfrac15$.

\vspace{2mm}

\noindent\udl{Level [0,2]}
\ban&&\back
\Gam(\emp,L_{-2})=
 \Gam(\emp,L_{-1})+\Di
\,,\quad
\Gam(\emp,L_{-1}^2)=
 \big[\Gam(\emp,L_{-1})+1\big]\Gam(\emp,L_{-1})
\,,\nt&&\back
\Gam(\emp,L_{-1}W_{-1})=
 \big[\Gam(\emp,L_{-1})+1\big]\Gam(\emp,W_{-1})
\,,\quad
\Gam(\emp,W_{-2})=
 \Gam(\emp,W_{-1})-\chi\Gam(\emp,L_{-1})
\,,\nt&&\back
\Gam(\emp,W_{-1}^2)=
\big[\Gam(\emp,W_{-1})-\chi\big]\Gam(\emp,W_{-1})
 +\tfrac92 D_0 \Gam(\emp,L_{-1})
\,.\ean


\noindent\udl{Level [3,0]}
\quad We don't write the explicit forms of
$\Gam(W_{-1}^3,\emp)$, since it is very cumbersome.
\ban&&\back
\Gam(L_{-3},\emp)=
 \Gam(L_{-2},\emp)+\Di
\,,\quad
\Gam(L_{-2}L_{-1},\emp)=
 \big[\Gam(L_{-2},\emp)+1\big]\Gam(L_{-1},\emp)
\,,\nt&&\back
\Gam(L_{-1}^3,\emp)=
 \big[\Gam(L_{-1},\emp)+2\big]\Gam(L_{-1}^2,\emp)
\,,\quad
\Gam(L_{-2}W_{-1},\emp)=
 \big[\Gam(L_{-2},\emp)+1\big]\Gam(W_{-1},\emp)
\,,\nt&&\back
\Gam(L_{-1}^2W_{-1},\emp)=
 \big[\Gam(L_{-1},\emp)+2\big]\Gam(L_{-1}W_{-1},\emp)
\,,\quad
\Gam(L_{-1}W_{-2},\emp)=
 \big[\Gam(L_{-1},\emp)+2\big]\Gam(W_{-2},\emp)
\,,\nt&&\back
\Gam(L_{-1}W_{-1}^2,\emp)=
 \big[\Gam(L_{-1},\emp)+2\big]\Gam(W_{-1}^2,\emp)
\,,\quad
\Gam(W_{-3},\emp)=
 \Gam(W_{-2},\emp)+\chi\Gam(L_{-2},\emp)-\tfrac12w_1
\,,\nt&&\back
\Gam(W_{-2}W_{-1},\emp)=
 \big[\Gam(W_{-2},\emp)+2\chi\big]\Gam(W_{-1},\emp)
 +\tfrac92 D_\infty\Gam(L_{-1},\emp)
\,.\ean

\noindent\udl{Level [2,1]}
\quad Similarly, we don't write the explicit forms of
$\Gam(W_{-1}^2,L_{-1})$ and $\Gam(W_{-1}^2,W_{-1})$.
\ban&&\back
\Gam(L_{-2},L_{-1})=
 \big[\Gam(L_{-2},\emp)-1\big]\Gam(\emp,L_{-1})
\,,\quad
\Gam(L_{-2},W_{-1})=
 \big[\Gam(L_{-2},\emp)-1\big]\Gam(\emp,W_{-1})
\,,\nt&&\back
\Gam(L_{-1}^2,L_{-1})=
 \Gam(L_{-1},\emp)\big[\Gam(L_{-1},L_{-1})+2\Do \big]
\,,\quad
\Gam(L_{-1}^2,W_{-1})=
 \Gam(L_{-1},\emp)\big[\Gam(L_{-1},W_{-1})+3\wo \big]
\,,\nt&&\back
\Gam(L_{-1}W_{-1},L_{-1})=
 \Gam(L_{-1},\emp)\Gam(W_{-1},L_{-1})+2\Do\Gam(W_{-1},\emp)
\,,\nt&&\back
\Gam(L_{-1}W_{-1},W_{-1})=
 \Gam(L_{-1},\emp)\Gam(W_{-1},W_{-1})
 +3w_0\Gam(W_{-1},\emp)
 -(\Dinf-\Do)(\Dinf+1)\left(\tfrac92 D_1-2\chi^2\right)
\,,\nt&&\back
\Gam(W_{-2},L_{-1})=
 \Gam(W_{-2},\emp)\Gam(\emp,L_{-1})-2(-\winf+\wi+\wo)
\,,\nt&&\back
\Gam(W_{-2},W_{-1})=
 \big[\Gam(W_{-2},\emp)-2\chi\big]\Gam(\emp,W_{-1})
 -\tfrac92 D_0\Gam(\emp,L_{-1})
\,,\ean
where $D_1:=\kappa(\Delta_1+\tfrac15)-\tfrac15$.

\vspace{2mm}

\noindent\udl{Level [1,2]}
\quad Similarly, we don't write the explicit forms of
$\Gam(L_{-1},W_{-1}^2)$ and $\Gam(W_{-1},W_{-1}^2)$.
\ban&&\back
\Gam(L_{-1},L_{-2})=
 \Gam(L_{-1},\emp)\big[\Gam(\emp,L_{-2})-1\big]
\,,\quad
\Gam(W_{-1},L_{-2})=
 \Gam(W_{-1},\emp)\big[\Gam(\emp,L_{-2})-1\big]
\,,\nt&&\back
\Gam(L_{-1},L_{-1}^2)=
 \big[\Gam(L_{-1},L_{-1})+2\Dinf \big] \Gam(\emp,L_{-1})
\,,\quad
\Gam(W_{-1},L_{-1}^2)=
 \big[\Gam(W_{-1},L_{-1})+3\winf \big] \Gam(\emp,L_{-1})
\,,\nt&&\back
\Gam(L_{-1},L_{-1}W_{-1})=
 \Gam(L_{-1},W_{-1})\Gam(\emp,L_{-1})+2\Dinf\Gam(\emp,W_{-1})
\,,\nt&&\back
\Gam(W_{-1},L_{-1}W_{-1})=
 \Gam(W_{-1},W_{-1})\Gam(\emp,L_{-1})+3\winf\Gam(\emp,W_{-1})
\,,\nt&&\back
\Gam(L_{-1},W_{-2})=
 \Gam(L_{-1},\emp)\Gam(\emp,W_{-2})+2(-\winf+\wi+\wo)
\,,\nt&&\back
\Gam(W_{-1},W_{-2})=
 \Gam(W_{-1},\emp)\big[\Gam(\emp,W_{-2})+2\chi\big]
 +\tfrac{9}{2}D_\infty\big[\Gam(\emp,L_{-1})-2\Di\big]
 -\left(\tfrac92D_1-2\chi^2\right)(-3\Dinf+\Di+3\Do)
\,.\ean


\noindent\udl{Level [0,3]}
\quad Similarly, we don't write the explicit form of $\Gam(\emp,W_{-1}^3)$.
\ban&&\back
\Gam(\emp,L_{-3})=
 \Gam(\emp,L_{-2})+\Di
\,,\quad
\Gam(\emp,L_{-2}L_{-1})=
 \big[\Gam(\emp,L_{-2})+1\big]\Gam(\emp,L_{-1})
\,,\nt&&\back
\Gam(\emp,L_{-1}^3)=
 \big[\Gam(\emp,L_{-1})+2\big]\Gam(\emp,L_{-1}^2)
\,,\quad
\Gam(\emp,L_{-2}W_{-1})=
 \big[\Gam(\emp,L_{-2})+1\big]\Gam(\emp,W_{-1})
\,,\nt&&\back
\Gam(\emp,L_{-1}^2W_{-1})=
 \big[\Gam(\emp,L_{-1})+2\big]\Gam(\emp,L_{-1}W_{-1})
\,,\quad
\Gam(\emp,L_{-1}W_{-2})=
 \big[\Gam(\emp,L_{-1})+2\big]\Gam(\emp,W_{-2})
\,,\nt&&\back
\Gam(\emp,L_{-1}W_{-1}^2)=
 \big[\Gam(\emp,L_{-1})+2\big]\Gam(\emp,W_{-1}^2)
\,,\quad
\Gam(\emp,W_{-3})=
 \Gam(\emp,W_{-2})-\chi\Gam(\emp,L_{-2})+\tfrac12\wi
\,,\nt&&\back
\Gam(\emp,W_{-2}W_{-1})=
 \big[\Gam(\emp,W_{-2})-2\chi\big]\Gam(\emp,W_{-1})
 +\tfrac92 D_0\Gam(\emp,L_{-1})
\,.\ean

}

\section{Constraint on 3-point function which contains two degenerate operators}
\label{s:3pt}

In this appendix, we study some properties of 3-point functions
$\langle \Delta_2,w_2|\phi_{\Delta_3,w_3}|\Delta_1, w_1\rangle$ where two
operators have level-1 singular vectors.  It may help
us to remove the difficulty for general quiver in \S\ref{SU3SU2}
where the 3-point functions (\ref{C1})--(\ref{C2}) vanishes automatically
if the two of the vertex operators have the level-1 null state.
We show here that the conformal Ward identity implies much weaker
condition for the third vertex operator to have nonvanishing 3-point function.
It suggests that we need to modify these formulae for such special case.

We assume the bra and ket state to have such singular vectors:
\ba\label{lv1sing}
 (W_{-1} -\frac{3w_1}{2\Delta_1} L_{-1})|\Delta_1, w_1\rangle =0\,,\quad
 \langle \Delta_2, w_2|(W_{1} -\frac{3w_2}{2\Delta_2} L_{1})=0\,.
\ea
What we study in the following is the consequence of Ward identity
explained in eq.\,(\ref{sec:3pt}).  In order to make the explanation clearer,
it will be useful to introduce some notation  in \cite{Bowcock:1993wq}.

We start from the action of $L_n$, $W_n$ on primary field
\ba
\left[L_n,\phi_{\Delta,w}(z)\right]&=& z^{n+1}\partial\phi_{\Delta,w} + \Delta(n+1)
z^n\phi_{\Delta,w}(z)\,,\\
\left[W_n, \phi_{\Delta,w}(z)\right]&=& z^n\left\{
\frac{w}{2} (n+1)(n+2) +(n+2) z \hat W_{-1} + z^2 \hat W_{-2}\right\}
\phi_{\Delta,w}(z)\,.
\nn\ea
We define the operators as
\ba
e_n(z) &:=& L_n -2z L_{n-1}+  z^2L_{n-2}\nt
f_n(z) &=& W_n -3z W_{n-1} +3z^2 W_{n-2} -z^3 W_{n-3}\,,
\ea
which satisfy
\ba
[e_n(z), \phi_{\Delta,w}(z)]=[f_n(z), \phi_{\Delta,w}(z)]=0\,.
\ea
By combining them with highest weight condition, we find
\ba
&&\langle \Delta_2, w_2|\phi_{\Delta_3, w_3}(z) e_n(z) =
\langle \Delta_2, w_2|\phi_{\Delta_3, w_3}(z) f_n(z)=0\quad (\mbox{for } n<0)\,,
\quad~~\nt
&&\langle \Delta_2, w_2|\phi_{\Delta_3, w_3}(z) (e_0(z)-\Delta_2)=
\langle \Delta_2, w_2|\phi_{\Delta_3, w_3}(z) (f_0(z)-w_2)=0\,.
\quad~~
\ea

Suppose the bra state $\langle \Delta_2, w_2|$ has also the level-1 null state
as eq.\,(\ref{lv1sing}), then it gives a constraint on the operator $\phi_{\Delta_3,w_3}$
to have a nonvanishing 3-point function.
To derive it,  we rewrite the level-1 null state condition for
$\phi_{\Delta_2, w_2}$ in eq.\,(\ref{lv1sing}) as
\ba
\langle \Delta_2, w_2|\phi_{\Delta_3, w_3}\,(f_{1} - \frac{3w_2}{2\Delta_2} e_{1})=0 \,,
\ea
In the following we put $z=1$ and omit the argument of field in the following.
The action of $f_{1} - \frac{3w_2}{2\Delta_2} e_{1}$ on the ket vector
$|\Delta_1, w_1\rangle$
is evaluated by
\ba
(f_{1} - \frac{3w_2}{2\Delta_2} e_{1})|\Delta_1, w_1\rangle
=\left\{
-3(w_1-\frac{w_2 \Delta_1}{\Delta_2}) +\frac{3}{2}(\frac{3w_1}{\Delta_1}
-\frac{w_2}{\Delta_2})L_{-1} -W_{-2}
\right\}|\Delta_1, w_1\rangle\,.
\ea
The 3-point functions for the descendants are given as
\ba
&&\frac{\langle \Delta_2, w_2|\phi_{\Delta_3,w_3} L_{-1}|\Delta_1, w_1\rangle}{
\langle \Delta_2, w_2|\phi_{\Delta_3,w_3}|\Delta_1, w_1\rangle} =\Delta_1-\Delta_2+\Delta_3,\\
&&\frac{\langle \Delta_2, w_2|\phi_{\Delta_3,w_3} W_{-2}|\Delta_1, w_1\rangle}{
\langle \Delta_2, w_2|\phi_{\Delta_3,w_3}|\Delta_1, w_1\rangle} =
2w_1+w_2-w_3 -3 w_1\frac{\Delta_2-\Delta_3}{\Delta_1}\,.
\ea
By the requirement
$\langle \Delta_2, w_2|\phi_{\Delta_3, w_3}\,(f_{1} - \frac{3w_2}{2\Delta_2}
e_{1})|\Delta_1, w_2\rangle=0$,
we obtain the constraint on the third vertex
\ba\label{w3r}
w_3=\frac32(\Delta_1+\Delta_2-\Delta_3)\left(\frac{w_1}{\Delta_1}-\frac{w_2}{\Delta_2}\right)-w_1+w_2
\,.\ea

As we noted in \S\ref{SU3SU2}, the 3-point coefficients
in eqs.\,(\ref{C1})--(\ref{C2}) vanish when two of the vertex operators to have
level-1 null state.  So the relation which we obtained here is much weaker.
Conformal symmetry requires only
one linear relation between $\Delta_3$ and  $w_3$ to
have nonvanishing 3-point function.

It may be interesting to learn the implication of the formula (\ref{w3r}).
One possibility may be the third vertex is forced to have level-1
null state as the other two. Since such conclusion would have a serious
consequence in AGT-W relation, let us work more on it.

In order to have level-1 singular vector, we need a constraint
on $h,w$.  In order to see it, we need to impose
\ba
 L_1(W_{-1} -\frac{3w}{2\Delta} L_{-1})|\Delta, w\rangle =0 ,\quad
 W_1(W_{-1} -\frac{3w}{2\Delta} L_{-1})|\Delta, w\rangle =0.\label{cons}
\ea
After using the commutation relations
\ba
&&[L_1, L_{-1}]=2L_0\,, \quad
[L_1, W_{-1}]=[W_1, L_{-1}]=3W_0\,,\nt
&&\frac{2}{9} [W_1, W_{-1}]=\kappa \Lambda_0-\frac{1}{5}L_0\,,
\ea
where
$\kappa=\frac{32}{22+5c}$ and
$\Lambda_0=L_0^2+\frac{1}{5} L_0+\cdots$,
we find that the first equation in eq.\,(\ref{cons}) is satisfied automatically
but the second equation requires
\ba\label{wh}
\left(\frac{w}{\Delta}\right)^2=\kappa(\Delta+\frac{1}{5}) -\frac15\,.
\ea

For $c=2$ (namely $Q=0$), $\kappa$ becomes $1$.
So this constraint reduces to
\ba
w^2=\Delta^3\,.
\ea
For this case, we may parametrize
\ba
\Delta_a= p_a^2,\quad w_a=-p_a^3\quad (a=1,2)
\ea
and would like to see if similar condition exists for $\Delta_3,w_3$.
Putting this in eq.\,(\ref{w3r}) with $\Delta_3=p_3^2$, we obtain
\ba
w_3=\frac12 (p_1-p_2)(3p_3^2-(p_1-p_2)^2)\,.
\ea
Obviously it does not take the form of level-1 null state (which should be
equal to $-p_3^3$).
However, if we additionally put $p_3=p_2-p_1$, which may be regarded
as ``momentum conservation," we have
\ba
w_3=-(p_2-p_1)^3.
\ea
Such momentum conservation is needed for free field theory.
For such case, the third vertex operator needs to have the level-1
null state.  Otherwise, for Toda theory with exponential interaction,
the momentum conservation condition is in general broken
and the third vertex operator need not to have the level-1 null state.

\end{document}